\documentstyle[aps]{revtex}

\begin{document}

\title{Nucleon to Delta Weak Excitation Amplitudes in the Non-relativistic
Quark Model}
\author{J\`un L\'\i u, Nimai C. Mukhopadhyay and Lisheng Zhang  \\ 
{\it Physics Department, Rensselaer Polytechnic Institute, Troy, NY
12180-3590}}
\maketitle
\begin{abstract}
We investigate the nucleon to Delta(1232) vector and axial vector amplitudes
in the non-relativistic quark model of the Isgur-Karl variety. A particular
interest is to investigate the SU(6) symmetry breaking, due to color
hyperfine interaction. We compare the theoretical estimates to recent
experimental investigation of the Adler amplitudes by neutrino scattering.
\end{abstract}

\section{Introduction}

Recent claim \cite{1} by a group of experimentalists at Los Alamos that they
may be observing neutrino flavor oscillations is a dramatic example of high
topical interest in the low and medium energy neutrino physics. Such
experiments are now possible, for example, at Brookhaven, Fermilab and Los
Alamos, just to name a few laboratories among many facilities around the
world. As a by-product of this experimental opportunity of having excellent
medium energy neutrino beams ($E_\nu \sim 1-3\;GeV$) being readily
available, a rebirth of exploring with neutrinos the properties of hadrons,
particularly nucleons, both in their ground states and the resonance region,
is expected. There is a long history of such investigations [2-18] in the
Delta(1232) resonance region since the sixties, the most recent experiment
\cite{18} in the Delta region have been done at Brookhaven. Given the
topical interest of the structure of hadrons from the QCD point of view, the
exploration of nucleons and their excited states, {\it by both
electromagnetic and weak probes}, merit special attention. In this
exploration, the weak structure functions are difficult to determine, but
they provide valuable information, often complementary to that obtained by
the electromagnetic interaction.

Among all the excited states of the nucleon, the Delta(1232) is perhaps the
best studied one \cite{19,20}, by strong, electromagnetic and even weak
interactions, the last one being the focus of this paper. Along with the
nucleon, the Delta is of fundamental interest to quantum chromodynamics
(QCD) and its application to the problem of hadron structure. Until we learn
how to use QCD rigorously to compute hadron properties, particularly in its
low-energy (hence non-perturbative) domain, we must cope with models that
are ``QCD-inspired''. One of the most successful examples of these imperfect
constructs is the quark shell model (QM), in which the gluonic degrees of
freedoms are replaced by effective potentials. The origin of this can be
traced back to the pre-QCD sixties, when G\"ursey, Radicati \cite{20} and
Sakita \cite{21} described the nucleon and the Delta in the fundamental {\bf
56}-dimensional representation of the spin-flavor SU(6) symmetry group. In
this limit, the nucleon and the Delta wave functions are $|N>=|\{56\}^2S_s>$
and $|\Delta >=|\{56\}^4S_s>$ . They are degenerate in the symmetry limit.
The degeneracy is lifted by the color hyperfine interaction \cite{22}. Thus,
in the Isgur-Karl QM, the wave functions of the nucleon and the Delta are
\cite{23,24} :
\begin{equation}
|N>=a_s|^2S_s>+a_{s^{\prime }}|^2S_{s^{\prime
}}>+a_M|^2S_M>+a_D|^4D_M>+a_p|^2P_A>,
\end{equation}
\begin{equation}
|\Delta >=b_s|^4S_s>+b_{s^{\prime }}|^4S_{s^{\prime
}}>+b_D|^4D_s>+b_{D^{\prime }}|^2D_M>,
\end{equation}
where the a's and b's are determined by diagonalizing the QM Hamiltonian in
the N=2 harmonic oscillator basis.

The purpose of this work is to investigate the process \cite{2}
\begin{equation}
p+\nu _\mu \rightarrow \Delta ^{++}+\mu ^{-},
\end{equation}
in the framework of the non-relativistic QM, using wave functions (2).
Despite a long history of theoretical (Refs.[2-14]) and experimental [15-18]
investigations of the process, {\it no calculations of the amplitudes for
the weak transitions are available in the literature in the context of these
general wave-functions.} Our main objective in this paper is to remedy this.
We shall focus in this paper on the all relevant helicity amplitudes. There
are four transverse ones: $A_{1/2}^V,\;A_{3/2}^V\;$and $A_{1/2}^A,%
\;A_{3/2}^A $, where V stands for vector and A represents axial vector. We
shall also discuss the relevant longitudinal (scalar) weak amplitudes. From
the electromagnetic processes \cite{25}
\begin{equation}
N+\gamma \rightleftharpoons \Delta ,
\end{equation}
we know the vector helicity amplitudes $A_\alpha ^V$, both extracted from
the experiments and as computed in various theoretical approaches, such as
the quark model (QM), both non-relativistic \cite{23,24} and relativized
\cite{26}, bags \cite{27}, topological \cite{13} and non-topological \cite
{28} solitons, and by the lattice gauge theoretic method \cite{29}.
Similarly thorough theoretical investigations on the axial vector amplitudes
have been suggested in a recent investigation \cite{14} in the framework of
QM. This is our objective here: to compute the vector and axial vector
amplitudes in the framework of QM. We shall relate them to the recently
extracted Adler \cite{4} amplitudes from the Brookhaven neutrino experiment
\cite{18}, which has studied the reaction
\begin{equation}
\nu _\mu +d\rightarrow \mu ^{-}+\Delta ^{++}(1232)+n_s,
\end{equation}
$n_s$ being spectator neutron. When necessary, we shall go back to the older
experiments as well.

Our main goal is to investigate the success of the non-relativistic QM in
reproducing the measured transition form factors for the process (3). In the
case of (4), the phenomenologically extracted magnetic dipole ($M1$) and
electric quadrupole ($E2$) amplitudes\cite{25} at the real photon point are
considerably {\em larger} than the values obtained in the QM. Their values
away from the real photon point are not very well-known experimentally as
yet. This is going to change with the advent of the electron facility called
the CEBAF. The theoretical deficit of the transition magnetic amplitude,
compared to the observed one, seems to be confirmed by the low-energy
Compton scattering, where the resonant magnetic polarizability appears to be
completely dominated\cite{30} by the Delta contribution. One of our goals
here is to see if the axial-vector analogue of the $M1$ amplitude in the
process (3) can be reproduced in the non-relativistic QM. We shall also make
an estimate of the axial-vector analogue of the E2 amplitudes, which would
be zero in the SU(6) symmetry limit. Thus, its non-zero value would be a
direct manifestation of the effect of the color magnetism, just as a
non-zero value of the $E2$ amplitude is in the vector sector.

Remainder of this paper is organized as follows: We derive an effective
Hamiltonian for the QM with vector and axial vector interactions in section
II, wherein we also find a relation between transverse vector and axial
vector amplitudes. The calculation of helicity amplitudes is done in section
III. We find relations between the helicity amplitudes and Adler's
form-factors in section IV, for the comparison of our model estimates with
experiments. We collect some exact SU(6) relations in section V. Section VI
contains the main results of our QM calculation and its comparison with
experiments. In this section, we introduce an axial vector analogue of the $%
E2/M1$ ratio. The SU(6) breaking results in this ratio becoming non-zero.
Neutrino experiments can give us an estimate for this ratio. Our conclusions
are summarized in section VII.

\section{The quark transition Hamiltonian}

The electroweak interaction at the quark level is incorporated in the Dirac
equation in the usual fashion. We introduce the leptonic current $l_\mu $
for the weak process, which is the analogue of the electromagnetic vector
potential $A_\mu $:
\begin{equation}
\sum_{i=1}^3\left( p_{\mu ,i}\gamma ^\mu -m_i-\gamma ^\mu \left( 1-\gamma
_5\right) l_{\mu ,i}\right) u=0,
\end{equation}
so that the Hamiltonian is

\begin{equation}
H=\sum_{i=1}^3\left( {\bf \alpha }\cdot \left( {\bf p}_i-\left( 1-\gamma
_5\right) {\bf l}_i\right) +\beta m_i+\left( 1-\gamma _5\right)
l_{0,i}\right) .
\end{equation}
By doing a free Foldy-Wouthuysen reduction \cite{31}, the quark Hamiltonian
can be truncated to
\begin{equation}
H=\sum_{i=1}^3\left( m_i+\frac{p^2}{2m_i}\right) +H_{int}^V-H_{int}^A,
\end{equation}
with
\begin{equation}
H_{int}^V=a^V\sum_{i=1}^3\left( l_{0i}-\frac 1{2m_i}\left( {\bf p}_i\cdot
{\bf l}_i+{\bf l}_i\cdot {\bf p}_i\right) -\frac 1{2m_i}\sigma \cdot \left(
\nabla \times {\bf l}_i\right) +O\left( m_i^{-2}\right) \right) ,
\end{equation}
\begin{equation}
H_{int}^A=a^A\sum_{i=1}^3\left( -\sigma \cdot {\bf l}_i+\frac 1{2m_i}\sigma
\cdot \left( {\bf p}_i\,l_{0i}+l_{0i}{\bf p}_i\right) +O\left(
m_i^{-2}\right) \right) .
\end{equation}
Here $i$ is the quark index, $a^V$ and $a^A$ are factors, which, in general,
can be different. Unless otherwise stated, we shall take these factors to be
unity. We take quark masses $m_i$ to be the same, $m$, for up and down
quarks. We can see that the $H_{int}^V$ is the same as interaction
Hamiltonian of the electromagnetic interaction except for the obvious
difference in coupling constants. The lepton current $l_\mu $ can be
defined, in analogy to the electromagnetic vector potential $A_\mu $, as
follows:
\begin{equation}
{\bf l}=\sqrt{\frac{4\pi \alpha _W}{2K_0}}\sum_{k,\lambda }\left( {\bf %
\epsilon }_\lambda a_{k\lambda }e^{i{\bf k\cdot r}}+{\bf \epsilon }_\lambda
^{*}a_{k\lambda }^{+}e^{-i{\bf k\cdot r}}\right) ,
\end{equation}
\begin{equation}
l_0=\sqrt{\frac{4\pi \alpha _W}{2K_0}}\sum_{k,\lambda }\left( a_{k\lambda
}e^{i{\bf k\cdot r}}+a_{k\lambda }^{+}e^{-i{\bf k\cdot r}}\right) ,
\end{equation}
where we introduce a ``weak'' fine structure constant, $\alpha _W,$ for
emitting or absorbing a weak quantum, W, with four-momentum $q_\mu
\;(k_0;0,0,k{\bf )}$. Here $K_0$ is the energy transfer for $q_\mu q^\mu =0$%
, given by
\begin{equation}
K_0=\frac{M_\Delta ^2-M^2}{2M_\Delta }\approx 255.8MeV.
\end{equation}
This is the analogue of the real photon point. We use $K_0$ in the
normalization factor under the radical sign in Eq.(11,12), a practice
adopted by many \cite{26,32}. This factor cancels out in the form-factor
expression, having no influence on the form-factors we shall calculate. For
the transverse amplitude, we use the boson polarization vector to be ${\bf %
\epsilon }=-\frac 1{\sqrt{2}}\left( 1,i,0\right) $. The vector transverse
Hamiltonian can be separated into two pieces:
\begin{equation}
H_{int,\;trans.}^V=H_I^V+H_{II}^V,
\end{equation}
with
\begin{equation}
H_I^V=\sqrt{\frac{4\pi \alpha _W}{2K_0}}\frac k{m\sqrt{2}}\left( 3\frac{%
(\tau _3)^{(3)}}2\right) \left( s_x^{(3)}+is_y^{(3)}\right) \exp \left( -i%
\sqrt{\frac 23}k\lambda _z\right) ,
\end{equation}
\begin{equation}
H_{II}^V=\sqrt{\frac{4\pi \alpha _W}{2K_0}}\frac 1{m\sqrt{2}}\left( 3\frac{%
(\tau _3)^{(3)}}2\right) \sqrt{\frac 23}\left( p_{\lambda
_x}^{(3)}+ip_{\lambda _y}^{(3)}\right) \exp \left( -i\sqrt{\frac 23}k\lambda
_z\right) ,
\end{equation}
taking $\frac{{\bf \sigma }}2={\bf s}.$

The correspondence between the usual photon case and the virtual W-boson
exchange discussed above, needed for the process (3), can be seen by the
appropriate identification of the photon variables to the W-boson variables.
For transverse axial transition, $l_0$ does not contribute, and the other
term is
\begin{equation}
H_{trans.}^A=\sqrt{\frac{4\pi \alpha }{2K_0}}\sqrt{2}\left( 3\frac{(\tau
_3)^{(3)}}2\right) \left( s_x^{(3)}+is_y^{(3)}\right) \exp \left( -i\sqrt{%
\frac 23}k\lambda _z\right) .
\end{equation}
Thus we have an important relation
\begin{equation}
H_{trans.}^A=\frac{2m}k\frac{e^A}{e^V}H_I^V.
\end{equation}
where $e^A/e^V$, denote the possible coupling difference between vector and
axial vector parts. If we compare directly with photon transitions, $e^V$ is
simply $e$.

\section{Calculation of the Helicity Amplitudes}

The calculation of the longitudinal and scalar amplitudes requires a
discussion of the possible lack of current conservation for chiral vector
and axial vector currents in the QM space chosen.. Though discussions on
this issue have been made from time to time in the literature \cite{8,32},
this issue is not settled in our opinion. In this paper, we simply avoid
tackling this issue. We begin here with a discussion of transverse helicity
amplitudes. The leading orders of these amplitudes do not suffer from the
uncertainties of the lack of current conservation.

The calculation of the transverse helicity amplitudes for the vector current
is standard \cite{23,24,32}. The wave functions of $\Delta $ and $N,$ taking
the effects of the color-hyperfine interaction, can be expressed in terms of
SU(6) basis functions as
\begin{equation}
|\Delta >=0.97|^4S_s>+0.20|^4S_s^{\prime }>-0.097|^4D_s>+0.065|^2D_M>,
\end{equation}
\begin{equation}
|N>=0.95|^2S_s>-0.24|^2S_s^{\prime }>-0.20|^2S_M>-0.042|^4D_M>,
\end{equation}
ignoring the tiny contribution from $|^2P_A>$ in (1). These two wave
functions correspond \cite{24} to energy levels $E_\Delta =1230MeV$, and $%
E_N=940MeV.\,$ The transverse helicity amplitudes for the electromagnetic
current are defined in the photon-nucleon CM frame \cite{24}:
\begin{equation}
A_\lambda ^V=<\Delta ,M_J=\lambda |H_{int}^V|\gamma N,\lambda =\lambda
_\gamma -\lambda _N,{\bf k}>,
\end{equation}
where the helicity $\lambda $ is 3/2 or 1/2. These equations are
straightforwardly generalizable to the case of our interest, introducing
virtual W boson mediating the charged weak current that produces the
reaction (3). From these quantities, we can define the familiar \cite{23,24}
transverse electromagnetic multipoles:
\begin{equation}
M1=-\frac 1{2\sqrt{3}}\left( 3A_{3/2}^V+\sqrt{3}A_{1/2}^V\right)
\end{equation}
\begin{equation}
E2=\frac 1{2\sqrt{3}}\left( A_{3/2}^V-\sqrt{3}A_{1/2}^V\right)
\end{equation}
We can also define the longitudinal and scalar amplitudes $L^V$ and S$^V$,
for which the virtual boson or photon helicity is zero. Clearly, these two
amplitudes are related \cite{32}:
\begin{equation}
k_0S^V=kL^V,
\end{equation}
a relation which can \cite{32} be violated due to truncation of the model
space in the QM. The relation (24) is a consequence of the conserved vector
current (CVC), or equivalently, the gauge invariance of the vector current.

The calculation of the transverse, longitudinal (and scalar) amplitudes for
the vector current have been done by many authors \cite{23,24,32}. Thus we
do not discuss them here. We shall give here a brief discussion for the
axial vector amplitudes. The longitudinal axial transition operator is given
by
\begin{equation}
H_{int}^{AL}=a\sum_{i=1}^3-\sigma _{iz}l_z,
\end{equation}
where we are considering the axial vector term only, Eq.(10). Following
(15), we have
\begin{equation}
H_{int}^{AL}=\sqrt{\frac{4\pi \alpha _W}{2K_0}}\left( -3\sigma
_z^{(3)}\right) \exp \left( -i\sqrt{\frac 23}k\lambda _z\right) .
\end{equation}
 From Eq.(10), the scalar term is
\begin{equation}
H_{int}^{AS}=a\sum_{i=1}^3\frac 1{2m_i}\left\{ {\bf \sigma }_i\cdot {\bf p}%
_i,\,l_{0i}\right\} .
\end{equation}
Following (16), we have
\begin{equation}
\begin{array}{c}
H_{int}^{AS}=
\sqrt{\frac{4\pi \alpha _W}{2K_0}} \\ \times \left( \frac 1{m\sqrt{3}}\left(
p_{\lambda _{+}}\sigma _{-}^{(3)}+p_{\lambda _{-}}\sigma _{+}^{(3)}\right)
-\frac km\left[ \frac{n_\Delta -n_N}{\overline{k}^2}-\frac 16\right]
\sqrt{2}%
\sigma _z^{(3)}\right) \exp \left( -i\sqrt{\frac 23}k\lambda _z\right) ,
\end{array}
\end{equation}
where $\overline{k}=k/\alpha _{HO}$, $\alpha _{HO}$ being the harmonic
oscillator parameter of the IK model, which has a value \cite{23} of 320MeV
from the fitting of the nucleon spectra. Here $n_\Delta $ and $n_N$ are the
principal quantum numbers of $\Delta $ and nucleon. The longitudinal and
scalar amplitudes are defined respectively as
\begin{equation}
L^A=<\Delta ;J_z=\frac 12|H_{int}^{AL}|N;J_z=\frac 12>,
\end{equation}
\begin{equation}
S^A=-<\Delta ;J_z=\frac 12|H_{int}^{AS}|N;J_z=\frac 12>.
\end{equation}
These amplitudes are shown in Table \ref{key1}.

We shall now use the computed values of $A_{3/2}^V$ and $\sqrt{3}A_{1/2}^V$
, obtained with $H_I^{(1)}$ of Eq.(14), and apply the relation (17) to
compute the $A_{3/2}^A$ and $\sqrt{3}A_{1/2}^A$. The results are shown in
Table \ref{key2}. In analogy to $M1$ and $E2$, discussed earlier
(Eqs.(22,23)), we can also compute the amplitudes $\left( M1\right) ^A$ and
$%
\left( E2\right) ^A$ , in the form:
\begin{equation}
\left( M1\right) ^A=A(\overline{k})f_M^A(\overline{k}),\;\left( E2\right)
^A=A(\overline{k})f_E^A(\overline{k}),
\end{equation}
Here
\begin{equation}
A(\overline{k})=\sqrt{\frac{4\pi \alpha _W}{2K_0}}e^{-\overline{k}^2/6},
\end{equation}
and $f_M^A(\overline{k})$, $f_E^A(\overline{k})$ are model-dependent, to be
explicitly given later.

QM calculations of $A_{1/2}^V$, $A_{3/2}^V$, $L^V$ and $S^V$ are standard in
the literature. For completeness, we have collected these results in our
notation in Table \ref{key3}. Bourdeau and Mukhopadhyay \cite{32} have
discussed the violation of (24) in the IK and other quark models. Numerical
results of the form-factors from IK model are shown in section VI,
subsection D. Ratios of helicity amplitudes are shown in Figs. 1 and 2.
Comparison with experiments is done through the Adler form-factors discussed
below.

\section{The Adler-Rarita-Schwinger formalism}

The standard method in the theoretical treatment of weak interaction process
in (3) follows the Rarita-Schwinger \cite{33,34} formalism and the $%
N\rightarrow \Delta $ transition form factors are introduced following a
notation due to Adler \cite{4} and Llewellyn Smith \cite{2}. Thus, the
invariant matrix element is
\begin{equation}
{\bf M}=<\mu ^{-}\Delta ^{++}|H_{int}|\nu p>=\frac{G_F\cos \theta
}{\sqrt{2}}%
j_\alpha <\Delta ^{++}|V^\alpha -A^\alpha |p>.
\end{equation}
where G$_F$ the Fermi constant, and $\theta $ is the Cabibbo angle,
\begin{equation}
j^\alpha =\overline{u}_\mu \gamma ^\alpha (1-\gamma ^5)u_\nu
\end{equation}
is the weak lepton current. We can decompose the invariant matrix element as
\cite{2,4,10}

$$
\frac{{\bf M}}{\sqrt{3}}=\frac G{\sqrt{2}}\overline{\psi }_\alpha \{\left(
\frac{C_3^V}M\gamma _\lambda +\frac{C_4^V}{M^2}(P_\Delta )_\lambda +\frac{%
C_5^V}{M^2}(P_p)_\lambda \right) \gamma _5F^{\lambda \alpha }+C_6^Vj^\alpha
\gamma _5
$$
\begin{equation}
+\left( \frac{C_3^A}M\gamma _\lambda +\frac{C_4^A}{M^2}(P_\Delta )_\lambda
\right) F^{\lambda \alpha }+C_5^Aj^\alpha +\frac{C_6^A}{M^2}(q)^\alpha
q^\lambda j_\lambda \}u,
\end{equation}
where
\begin{equation}
F^{\lambda \alpha }=q^\lambda j^\alpha -q^\alpha j^\lambda ,
\end{equation}
$P_\Delta $ is the Delta four-momentum, $\psi _\mu $ is the Delta vector
spinor and $u$, the proton spinor. The $C_i^{V,A}$ are the so-called Adler
form-factors \cite{4}. These form factors can be related to the helicity
amplitudes through calculation of ${\bf M}$ projected in different
polarizations. The factor $\sqrt{3}$ in Eq.(35) above comes from the isospin
relations, such as that between the weak and electromagnetic matrix
elements:
\begin{equation}
<\Delta ^{++}|V^\alpha |p>=\sqrt{3}<\Delta ^{+}|V_{EM}^\alpha |p>.
\end{equation}

After separating $j_\alpha $, we get%
$$
\frac 1{\sqrt{3}}<\Delta ^{++}|-A^\alpha |p>=\overline{\psi }_\alpha \left(
\frac{C_3^A}M\gamma _\lambda +\frac{C_4^A}{M^2}(P_\Delta )_\lambda \right)
q^\lambda u
$$
\begin{equation}
-\overline{\psi }_\lambda \left( \frac{C_3^A}M\gamma _\alpha +\frac{C_4^A}{%
M^2}(P_\Delta )_\alpha \right) q^\lambda u+\overline{\psi }_\alpha C_5^Au+%
\overline{\psi }_\lambda \frac{C_6^A}{M^2}q^\lambda q_\alpha u.
\end{equation}
Recent experiments on the process (3) at the Brookhaven National Laboratory
\cite{18} have been analyzed in terms of the Adler form factors. Thus, we
must find relations between the helicity amplitudes, discussed above, and
the Adler amplitudes.

In the $\Delta $ rest frame, which is also the $WN$(or $\gamma N$) CM frame,
\begin{equation}
(P_\Delta )_0=M_\Delta =E_N+k_0=\sqrt{M^2+{\bf k}^2}+k_{0,}
\end{equation}
${\bf p}_\Delta =0$ and $q^\lambda q_\lambda =k_0^2-{\bf k}^2$. We also
define the Delta vector spinors as
\begin{equation}
{\bf \psi }_{3/2}^\Delta ={\bf \epsilon }_{+}\chi _{+},
\end{equation}
\begin{equation}
{\bf \psi }_{1/2}^\Delta =\frac 1{\sqrt{3}}{\bf \epsilon }_{+}\chi _{-}+%
\sqrt{\frac 23}{\bf \epsilon }_0\chi _{+}.
\end{equation}
Here the polarization vectors are
\begin{equation}
{\bf \epsilon }_{+}=-\frac 1{\sqrt{2}}\left(
\begin{array}{c}
1 \\
i \\
0
\end{array}
\right) ,\;{\bf \epsilon }_{-}=\frac 1{\sqrt{2}}\left(
\begin{array}{c}
1 \\
-i \\
0
\end{array}
\right) ,\;{\bf \epsilon }_0=-\frac 1{\sqrt{2}}\left(
\begin{array}{c}
0 \\
0 \\
1
\end{array}
\right) .
\end{equation}
Thus, the vector spinor
\begin{equation}
\psi _{3/2}^\Delta =\epsilon _{+}\chi _{+}=\left\{ 0;-\frac 1{\sqrt{2}%
}\left(
\begin{array}{c}
\chi _{+} \\
0
\end{array}
\right) ,-\frac i{\sqrt{2}}\left(
\begin{array}{c}
\chi _{+} \\
0
\end{array}
\right) ,0\right\} .
\end{equation}
and so on.

The nucleon spinors are
\begin{equation}
u_{\pm }^N=\left(
\begin{array}{c}
\chi ^{\pm } \\
\frac{{\bf \sigma }\cdot {\bf p}_N}{E_N+M}\chi ^{\pm }
\end{array}
\right) =\left(
\begin{array}{c}
\chi ^{\pm } \\
-\frac{{\bf \sigma }\cdot {\bf k}}{E_N+M}\chi ^{\pm }
\end{array}
\right) .
\end{equation}
Thus we can derive the following relations:
\begin{equation}
\begin{array}{c}
-A_{3/2}^V=
\sqrt{\frac{4\pi \alpha _W}{2K_0}}\frac k{E_N+M} \\ \times \left[ \frac{%
E_N+M+k_0}MC_3^V+\frac{k_0M_\Delta }{M^2}C_4^V+\frac{k_0E_N+{\bf k}^2}{M^2}%
C_5^V+C_6^V\right] ,
\end{array}
\end{equation}
\begin{equation}
\begin{array}{c}
\sqrt{3}A_{1/2}^V=\sqrt{\frac{4\pi \alpha _W}{2K_0}}\frac k{E_N+M} \\ \times
\left[ \frac{-E_N-M+k_0}MC_3^V+\frac{k_0M_\Delta }{M^2}C_4^V+\frac{k_0E_N+%
{\bf k}^2}{M^2}C_5^V+C_6^V\right] ,
\end{array}
\end{equation}
\begin{equation}
\sqrt{\frac 32}L^V=\sqrt{\frac{4\pi \alpha _W}{2K_0}}\frac k{E_N+M}\left[
\frac{k_0}MC_3^V+\frac{k_0M_\Delta }{M^2}C_4^V+\frac{k_0E_N}{M^2}%
C_5^V+C_6^V\right] ,
\end{equation}
\begin{equation}
\sqrt{\frac 32}S^V=\sqrt{\frac{4\pi \alpha _W}{2K_0}}\frac k{E_N+M}\left[
\frac kMC_3^V+\frac{kM_\Delta }{M^2}C_4^V+\frac{kE_N}{M^2}C_5^V\right] ,
\end{equation}
\begin{equation}
-A_{3/2}^A=\sqrt{\frac{4\pi \alpha _W}{2K_0}}\left[ \left( k_0+\frac{{\bf k}%
^2}{E_N+M}\right) \frac 1MC_3^A+\frac{k_0M_\Delta }{M^2}C_4^A+C_5^A\right] ,
\end{equation}
\begin{equation}
-\sqrt{3}A_{1/2}^A=\sqrt{\frac{4\pi \alpha _W}{2K_0}}\left[ \left(
k_0-\frac{%
{\bf k}^2}{E_N+M}\right) \frac 1MC_3^A+\frac{k_0M_\Delta }{M^2}%
C_4^A+C_5^A\right] ,
\end{equation}
\begin{equation}
-\sqrt{\frac 32}L^A=\sqrt{\frac{4\pi \alpha _W}{2K_0}}\left[ \frac{k_0}%
MC_3^A+\frac{k_0M_\Delta }{M^2}C_4^A+C_5^A-\frac{{\bf k}^2}{M^2}C_6^A\right]
,
\end{equation}
\begin{equation}
-\sqrt{\frac 32}S^A=\sqrt{\frac{4\pi \alpha _W}{2K_0}}\left[ \frac kMC_3^A+%
\frac{kM_\Delta }{M^2}C_4^A-\frac{k_0k}{M^2}C_6^A\right] .
\end{equation}
 From the relations for the vector amplitudes, we get
\begin{equation}
A_{3/2}^V+\sqrt{3}A_{1/2}^V=-\sqrt{\frac{4\pi \alpha _W}{2K_0}}\frac{2k}%
MC_3^V,
\end{equation}
by far the {\em most important} vector contribution in the $N\rightarrow
\Delta $ transition. We also get
\begin{equation}
\frac 12\left( -A_{3/2}^V+\sqrt{3}A_{1/2}^V\right) -\sqrt{\frac
32}L^V=\sqrt{%
\frac{4\pi \alpha _W}{2K_0}}\frac{k^3}{M^2(E_N+M)}C_5^V,
\end{equation}
\begin{equation}
\sqrt{\frac 32}\left( kL^V-k_0S^V\right) =\sqrt{\frac{4\pi \alpha _W}{2K_0}}%
\left( E_N-M\right) C_6^V.
\end{equation}
By CVC or gauge invariance, the left-hand side is zero identically. Thus we
get a CVC relation, $C_6^V=0$, a relation {\em which will be mildly violated
by the IK quark model calculation}, due to its inability to make the
left-hand side of (55) vanish identically \cite{32}.

Rearranging the axial-vector transition helicity amplitudes, we have
\begin{equation}
\sqrt{\frac{4\pi \alpha _W}{2K_0}}C_3^A=\frac M{2\left( E_N-M\right) }\left(
\sqrt{3}A_{1/2}^A-A_{3/2}^A\right) .
\end{equation}
We can define relations analogous to (22) and (23):
\begin{equation}
\left( M1\right) ^A=-\frac 1{2\sqrt{3}}\left( 3A_{3/2}^A+\sqrt{3}%
A_{1/2}^A\right) ,
\end{equation}
\begin{equation}
\left( E2\right) ^A=\frac 1{2\sqrt{3}}\left( A_{3/2}^A-\sqrt{3}%
A_{1/2}^A\right) .
\end{equation}
We also have the relations
\begin{equation}
\sqrt{\frac{4\pi \alpha _W}{2K_0}}C_3^A=-\sqrt{3}\frac M{2\left(
E_N-M\right) }\left( E2\right) ^A,
\end{equation}
\begin{equation}
\sqrt{\frac{4\pi \alpha _W}{2K_0}}\left( C_5^A+\frac{M_\Delta k_0}%
MC_4^A\right) =-\frac 12\left( \sqrt{3}A_{1/2}^A+A_{3/2}^A\right)
-\frac{k_0}%
M\sqrt{\frac{4\pi \alpha _W}{2K_0}}C_3^A,
\end{equation}
\begin{equation}
=\frac{\sqrt{3}}2\left( \frac{k_0+E_N-M}{E_N-M}\left( E2\right) ^A+\left(
M1\right) ^A\right) .
\end{equation}

The amplitudes $S^A$ and $L^A$ are connected by the identity
\begin{equation}
\label{62}kL^A-k_0S^A=-\sqrt{\frac 23}k\sqrt{\frac{4\pi \alpha _W}{2K_0}}%
\left[ C_5^A+\frac{k_0^2-{\bf k}^2}{M^2}C_6^A\right] .
\end{equation}
We get the conservation of the axial current (CAC) in the chiral limit ($%
m_\pi \rightarrow 0$) in which this identity reduces to%
$$
\quad \quad kL^A-k_0S^A=0.\quad \quad \quad \quad \quad \quad \quad \quad
\quad \quad \quad \quad \quad \quad \quad \quad \quad \quad (\ref{62}%
^{\prime })
$$
We must examine, if the chiral symmetry breaking and the truncation of the
model space in the QM results in the violation of the last relation. The
answer we find below is in the affirmative.

Using (49-52) we can derive the following relations:
\begin{equation}
C_6^A=\frac{M^2}{{\bf k}^2}\sqrt{\frac{2K_0}{4\pi \alpha _W}}\left[ -\frac
12\left( \sqrt{3}A_{3/2}^A+A_{1/2}^A\right) +\sqrt{\frac 32}L^A\right] ,
\end{equation}
\begin{equation}
C_5^A=-\sqrt{\frac 32}\sqrt{\frac{2K_0}{4\pi \alpha _W}}\left(
L^A-\frac{k_0}%
kS^A\right) -\frac{k_0^2-{\bf k}^2}{M^2}C_6^A,
\end{equation}
\begin{equation}
C_4^A=\frac{M^2}{kM_\Delta }\left[ -\sqrt{\frac 32}\sqrt{\frac{2K_0}{4\pi
\alpha _W}}S^A-\frac kMC_3^A+\frac{k_0k}{M^2}C_6^A\right] .
\end{equation}
Thus, Eqs. (56,63-65) complete the expressions for the four Adler axial
nuclear to Delta form factors, in terms of the calculated helicity
amplitudes.

The partial conservation of axial current (PCAC) relation can be expressed
in another way, as discussed earlier by Schreiner and von Hippel \cite{10}.
 From the pion pole dominance of the divergence of the axial current, taken
between the nucleon and the Delta states, we get the induced pseudoscalar
term given by the pion pole, exactly parallel to the weak current between
nucleon states, wherein the pion pole term also yields the induced
pseudoscalar term \cite{10,14}. Thus,
\begin{equation}
\frac{C_6^A}{M^2}=\frac{g_\Delta f_\pi }{2\sqrt{3}M\left( m_\pi
^2-q^2\right) },
\end{equation}
at the pion pole. Here $g_\Delta $ is the $\Delta ^{++}\rightarrow p\pi ^{+}$
coupling constant, recently redetermined by Davidson {\it et al}. \cite{25}:
\begin{equation}
g_\Delta =28.6\pm 0.3\;.
\end{equation}
Actually, the determination of $g_\Delta $ depends on which method we use in
the analysis, thereby yielding a much bigger theoretical error than that is
given in (67). In Eq.(66), $f_\pi $ is the pion decay constant
\begin{equation}
f_\pi \approx 0.97m_\pi .
\end{equation}
Taking the limit of the divergence of the axial current as $m_\pi
^2\rightarrow 0$ and $q^2\rightarrow 0$, we get the {\em off-diagonal}
Goldberger-Treiman relation
\begin{equation}
C_5^A(0)=\frac{g_\Delta f_\pi }{2\sqrt{3}M}.
\end{equation}
This off-diagonal cousin of the well-known {\em diagonal} Goldberger-Treiman
relation has not been given much attention in the literature, except in the
recent work by Hemmert {\it et al}. \cite{14}. Using (69), the estimate of $%
C_5^A(0)$ is
\begin{equation}
C_5^A(0)\approx 1.2.
\end{equation}
The relations (62') and (69) should be {\em simultaneously satisfied} if
PCAC is to be valid.

We shall further discuss the subject of PCAC and its validity in the QM in
subsection VI.B. Form-factors calculated here are shown in Figs. 3-10.

\section{The exact SU(6) symmetry relations}

Let us first start with the vector amplitudes. In the SU(6) symmetry limit
\cite{35},
\begin{equation}
M1\neq 0,\;E2=0.
\end{equation}
The Fermi-Watson theorem implies these multipoles to be purely imaginary on
top of the Delta resonance. Thus,
\begin{equation}
Im(E2)/Im(M1)\equiv EMR=0.
\end{equation}
In the QM, all amplitudes are purely real, hence E2/M1=0 in the SU(6) limit,
giving
\begin{equation}
A_{3/2}^V(SU(6))=\sqrt{3}A_{1/2}^V(SU(6)).
\end{equation}
Using this relation in Eq.(53), we get, in the SU(6) symmetry limit, the
largest vector form factor, C$_3^V$, given by (Table \ref{key2})
\begin{equation}
C_3^V(SU(6))=-\left( \sqrt{\frac{4\pi \alpha _W}{2K_0}}\frac kM\right)
^{-1}A_{3/2}^V(SU(6))=\frac M{\sqrt{3}m}e^{-\overline{k}^2/6}.
\end{equation}
Also, the left-hand side of (54) vanishes in the SU(6) limit, giving
\begin{equation}
C_5^V(SU(6))=0.
\end{equation}
Likewise, we can get
\begin{equation}
C_4^V(SU(6))=-\frac M{M_\Delta }C_3^V(SU(6)).
\end{equation}
Finally, the longitudinal vector response vanishes in the SU(6) limit:
\begin{equation}
L^V(SU(6))=S^V(SU(6))=0.
\end{equation}
Thus, the only non-vanishing multipole vector amplitude in the SU(6) limit
is the magnetic dipole amplitude, given by
\begin{equation}
M1(SU(6))=-2A_{1/2}^V(SU(6))=\frac{2k}{3m}A(\overline{k}).
\end{equation}
We can similarly discuss the axial vector amplitude in the SU(6) limit. The
only non-vanishing amplitude is
\begin{equation}
\left( M1\right) ^A(SU(6))=\frac 43A(\overline{k}),
\end{equation}
and
\begin{equation}
\left( E2\right) ^A(SU(6))=0.
\end{equation}
The axial vector transverse helicity amplitudes are:%
$$
\frac 1{\sqrt{3}}A_{3/2}^A(SU(6))=A_{1/2}^A(SU(6))=-\frac 23A(\overline{k}).
$$
The scalar and longitudinal axial vector amplitudes are:
\begin{equation}
S^A(SU(6))=\frac{\sqrt{2}}9\frac kmA(\overline{k}),
\end{equation}
\begin{equation}
L^A(SU(6))=-\frac{2\sqrt{2}}3A(\overline{k}).
\end{equation}
The Adler form factors become
\begin{equation}
C_3^A(SU(6))=C_6^A(SU(6))=0,
\end{equation}
\begin{equation}
C_5^A(SU(6))=\left( \frac 2{\sqrt{3}}+\frac 1{3\sqrt{3}}\frac{k_0}m\right)
e^{-\overline{k}^2/6},
\end{equation}
\begin{equation}
C_4^A(SU(6))=-\frac 1{3\sqrt{3}}\frac{M^2}{M_\Delta m}e^{-\overline{k}^2/6}.
\end{equation}
In Figs.1 and 2, we plot the ratios
\begin{equation}
r^b=\frac{A_{3/2}^b}{\sqrt{3}A_{1/2}^b}
\end{equation}
as functions of $Q^2=-q^2$, b=V, A, for the IK wave functions of the nucleon
and the Delta, and the wave-functions \cite{32} inspired by a model of Vento
{\it et al. }(VBJ){\it \ }\cite{36}. In the SU(6) limit, this ratio should
be unity for both V and A currents. The deviation from unity is due to the
SU(6) breaking interactions. In the IK case, that is from the color
hyperfine interaction.

The SU(6) relations described above are only approximate, since the SU(6)
symmetry is broken by the color hyperfine interaction. Thus, the IK model
wave functions and the actual experiments would violate these relations. The
degree of these violations is an interesting question and lots of
theoretical \cite{25,36,37} and experimental \cite{38} attention are being
given to it, since the finding of Davidson {\it et al. }\cite{25} that the
EMR is not zero from the existing old electromagnetic data. We hope our
study here of the axial vector amplitudes would trigger similar interest in
the {\em weak sector}.

We note that the SU(6) symmetry limit to the nucleon and the Delta wave
function provides consistency with the requirements of CVC. Thus, the CVC
requirement of current conservation is trivially satisfied. The vanishing of
$C_6^V$, required by the CVC limit, is again trivially true.

The CAC relation (62') is {\em not} satisfied in the SU(6) limit. However,
we get good agreement with the off-diagonal Goldberger-Treiman relation,
Eq.(69). Thus, going to the ``real photon'' point $k=k_0$, we can evaluate
the $C_5^A(Q^2=0)$ by using (84). We get
\begin{equation}
C_5^A(Q^2=0)(SU(6))=1.17,
\end{equation}
compared with the off-diagonal Goldberger-Treiman value of 1.2, a good
agreement. We can also see that the identity (62) is satisfied in the SU(6)
limit of QM, as it must. However, both sides are quite large at the real
photon point:
\begin{equation}
\frac{\sqrt{2}}3K_0\sqrt{\frac{4\pi \alpha _W}{2K_0}}e^{-\overline{K}%
_0^2/6}\left( \frac{K_0}{3m}+2\right) ,
\end{equation}
instead of the PCAC expectation of both-sides of (62) vanishing at the
chiral limit, Eq.(62').

\section{Main results of the quark model and comparison with experimental
results}

\subsection{Comparison with other models}

An extensive review of the old QM calculations by many authors has been
done by Schreiner and von Hippel \cite{10}, who have provided a detailed
comparison between theory and experiment, as available till 1973. The
readers are referred to their paper for a discussion. In Table \ref{key4},
we summarize the predictions of various form factors, vector and axial
vector, in different models, and compare them with the new QM results
reported here, as well as the recent estimates of Hemmert {\it et al.}\cite
{14}.

\subsection{CVC and PCAC}

We have seen earlier that the quantity $\left( kL^V-k_0S^V\right) $, the
left-hand side of Eq. (55), vanishes in the SU(6) limit, thereby satisfying
the CVC. Other quark model wave functions for the nucleon and the Delta,
such as those of the Isgur-Karl model and the VBJ model \cite{36}, discussed
below, do not satisfy this CVC constraint \cite{32}. The CVC violations in
the IK and VBJ models are demonstrated in Figs. 5 and 6 where the $C_5^V$
and the $C_6^V$ are plotted as functions of $-q^2$ for W=1230MeV and
W=1234MeV.

As noted earlier, the PCAC pion pole estimate of $C_6^A(q^2),$ Eq.(66), is
not fulfilled in the QM. The IK quark model, for example, produces a value
of $C_6^A(q^2)$ {\em much smaller} than what Eq.(66) suggests. This failure
is readily understandable, because the inclusion of the pion-pole term in a
QM that involves only quark degrees of freedom {\em and no mesons}, is not
legitimate. However, what is surprising is that the {\em off-diagonal
Goldberger-Treiman relation} (Eq.(69)) {\em is well-satisfied in the SU(6)
limit and by the IK quark model wave functions.} Thus, in the SU(6) limit,
we have Eq.(87); in the IK model, we get {\em \ }
\begin{equation}
C_5^A(0)(IK)=1.16.
\end{equation}
This, however, occurs due to an accident. In the chiral limit
\begin{equation}
\left( L^A-\frac{k_0}kS^A\right) _{chiral}\rightarrow 0,
\end{equation}
in Eq.(64), and $C_6^A$ would pick up a PCAC contribution, not present in
our QM calculations. The latter does not satisfy the chiral axial current
conservation and does not pick up the PCAC contribution in $C_6^A$. However,
these two theoretical inaccuracies somehow add up to the PCAC estimate in
our SU(6) and IK quark model wave functions. Further theoretical work is
needed to illuminate the nature of this happy accident.

While the degree of violation of CVC and PCAC should be as small as possible
in the quark model of excellent quality, this does not mean we prefer the
SU(6) limit to the IK model. That is because the color hyperfine interaction
and other dynamical considerations make the SU(6) limit inaccurate for
baryon spectroscopy. We must, therefore, search for a model beyond that
limit. Between the IK model and the VBJ models, the former is clearly
superior, as the violation of CVC is less in the former than in the latter,
and the IK model account for the baryon spectroscopy, while the VBJ model is
not so ambitious.

\subsection{SU(6) breaking}

We can go back to Figs. 1 and 2 to demonstrate the SU(6) breakings in the IK
\cite{23,24} and VBJ \cite{36} models. These effects are significant, though
difficult to measure experimentally. In Figs. 3 and 9, the SU(6) limits
coincide with the theoretical predictions of CVC and off-diagonal
Goldberger-Treiman relation. Here, more realistic quark models produce mild
violation of these important constraints. Such violation can arise from the
model truncation effects. Some authors have tried to remedy these with the
introduction of the form factors at the quark level \cite{8,11}. We shall
not do that, as we believe this is not a satisfactory remedy in the spirit
of QCD.

\subsection{Comparison with experimental results}

The experiments on neutrino scattering in the Delta resonance region are
very difficult, nevertheless they have been done in a number of labs, ANL,
CERN, Fermilab and BNL [15-18]. These experiments deal with single pion
production in the charged current reaction in hydrogen and deuterium. In
view of the extensive literature on the older experiments, we shall focus
here on the most recent one.

Latest experimental results \cite{18} come from the Brookhaven National
Laboratory(BNL). The analysis by Kitagaki {\it et al}. \cite{18}. makes some
strong assumptions, including the neglect of the background contributions
and use of the polynomial forms for the transition form factors, prescribed
by Adler \cite{4}. The dipole form they use is
\begin{equation}
F_V(Q^2)=\lambda (Q^2)/(1+Q^2/M_V^2)^2
\end{equation}
where $\lambda (Q^2)$ is a function introduced by Olsson {\it et al}. \cite
{39}:
\begin{equation}
\lambda ^2(Q^2)=1-\left( 0.053+0.017Q\right) \sin \left(
\frac{4.00Q}{1+0.22Q%
}\right) .
\end{equation}
Kitagaki {\it et al. }\cite{18} find, from a fit to their new data,
\begin{equation}
M_V=0.89_{-0.07}^{+0.04}\,GeV.
\end{equation}
We demonstrate the results from Kitagaki {\it et al.} by displaying $%
C_3^V(Q^2)$, using Adler parameters \cite{4}, in Fig. 3 as a function of Q$%
^2 $. We also plot the earlier experimental fit by Dufner and Tsai \cite{40}%
, who got a good fit to the data on the electroproduction of pions in the
Delta region, by using the form
\begin{equation}
|C_3^V|^2=(2.05)^2\left( 1+9(Q^2)^{1/2}\right) \exp \left(
-6.3(Q^2)^{1/3}\right) .
\end{equation}
We shall also use the simpler dipole form \cite{10}, which fits the above
data just as well:
\begin{equation}
C_3^V=2.05\left( 1+\frac{Q^2}{0.54}\right) ^{-2}.
\end{equation}
In the formulae above, $Q^2$ is in GeV$^2$. We notice that the IK model \cite
{23,24} does a reasonable job of describing the experimental data on $C_3^V$%
, except at $Q^2=0$, where it falls short (Fig. 3) in comparison to
experiment. This is not surprising, as we have encountered this deficit
already in the photoproduction of pions \cite{25}.

We now come to the axial-vector form factors. Kitagaki {\it et al}. \cite{18}
assume the Adler form for the dependence on $Q^2$, and then try to fit their
new data to a range of $M_A$. Thus, they use
\begin{equation}
C_i^A(Q^2)=\frac{c_i(0)\left( 1+a_iQ^2/(b_i+Q^2)\right) }{\left(
1+Q^2/M_A^2\right) ^2},\,\,i=3,4,5,
\end{equation}
where a's b's and c's are all model-dependent parameters determined in the
Adler model \cite{4}:
\begin{equation}
\begin{array}{c}
c_3(0)=a_3=b_3=0, \\
c_4(0)=-0.3, \\
c_5(0)=1.2, \\
a_4=a_5=-1.21, \\
b_4=b_5=2.
\end{array}
\end{equation}
By fitting $M_V$ and $M_A$ simultaneously to their data, Kitagaki {\it et
al}%
. get, with $M_V$ given Eq.(93):
\begin{equation}
M_A=0.97_{-0.11}^{+0.14}\,GeV.
\end{equation}
Their results are found to be consistent with other earlier experiments and
analysis. We thus compare our QM results with the above parameterization,
where the experimental information controls the value of $M_A$. Our
agreement with the empirical forms needed by the data of Kitagaki {\it et
al.%
} is satisfactory in the IK model (Figs. 3, 9).

Given our theoretical difficulties in computing the longitudinal structure
functions in the QM, we shall compare our calculations of the axial-vector
helicity amplitudes, $A_{3/2}^A$ and $A_{1/2}^A$, with the experiments via
the form factors $C_3^A,C_4^A,C_5^A$. In Figs. 7-9, we display this
comparison.

Some cautionary remarks are in order regarding the quality of comparison of
the IK (or any other) quark model and the ``experimental values'' of the
Adler form factors \cite{4} that we have displayed in Figs. 3-10. The
analysis of the latest experimental results on neutrino scattering by
Kitagaki {\it et al. }\cite{18} does not separate the individual form
factors directly by experiment. Also, it makes strong assumptions on the $%
Q^2 $ dependence of the nucleon the Delta form factors, basically following
the polynomial forms prescribed by Adler \cite{4}. Finally, it neglects the
non-resonant Born contributions \cite{25}, which may be small, but
uninvestigated at this time.

Below we give numerical expressions of the helicity amplitudes and the Adler
amplitudes in the IK model. Since the differences between $A_{3/2}^{V,A}$
and $\sqrt{3}A_{3/2}^{V,A}$ are small, we give $M1$ and $\left( M1\right) ^A$
instead:
\begin{equation}
M1=\frac 1{\sqrt{3}}\frac k{2m}A(\overline{k})\left(
2.0237+0.0623\overline{k%
}^2-0.00133\overline{k}^4\right) ,
\end{equation}
\begin{equation}
E2=-\frac 1{\sqrt{3}}\frac k{2m}A(\overline{k})\left( 6.26+0.627\overline{k}%
^2+0.252\overline{k}^4\right) \times 10^{-3},
\end{equation}
\begin{equation}
S^V=-\frac 1{6\sqrt{15}}A(\overline{k})\overline{k}^2\left( 0.0996+0.00133%
\overline{k}^2\right) ,
\end{equation}
\begin{equation}
L^V=\sqrt{3}A(\overline{k})\frac km\left( -0.00104+0.000411\overline{k}%
^2\right) ,
\end{equation}
\begin{equation}
\left( M1\right) ^A=\frac 1{\sqrt{3}}A(\overline{k})\left( 2.0204+0.0625%
\overline{k}^2-0.00133\overline{k}^4\right) ,
\end{equation}
\begin{equation}
\left( E2\right) ^A=\frac 1{\sqrt{3}}A(\overline{k})\left( 0.325\overline{k}%
^2-0.251\overline{k}^4\right) \times 10^{-3},
\end{equation}
\begin{equation}
L^A=-A(\overline{k})\left( 0.8235+0.0225\overline{k}^2-0.000935\overline{k}%
^4\right) ,
\end{equation}
\begin{equation}
S^A=\sqrt{\frac 23}\frac kmA(\overline{k})\left( 0.338+0.00546\overline{k}%
^2-0.000205\overline{k}^4\right) .
\end{equation}

The IK model parameters have some uncertainties due to variable estimates in
the baryon eigenvalues. So we present below the result of our calculation,
using another set of IK model parameters, given by Gershtein and Dzhikiya
\cite{24}. The wave-functions in this case are given by
\begin{equation}
|N>=0.96|^2S_s>-0.18|^2S_{s^{\prime
}}>-0.22|^2S_M>-0.051|^4D_M>-0.0039|^2P_A>,
\end{equation}
\begin{equation}
|\Delta >=0.98|^4S_s>+0.16|^4S_{s^{\prime }}>-0.11|^4D_s>+0.088|^2D_M>.
\end{equation}
They correspond to energy values of $M_N=944MeV$ and $M_\Delta =1234MeV.$
The weak amplitudes are:
\begin{equation}
M1=\frac 1{\sqrt{3}}\frac k{2m}A(\overline{k})\left(
2.1159+0.0564\overline{k%
}^2-0.00084\overline{k}^4\right) ,
\end{equation}
\begin{equation}
E2=-\frac 1{\sqrt{3}}\frac k{2m}A(\overline{k})\left( 10.29+1.03\overline{k}%
^2+0.236\overline{k}^4\right) \times 10^{-3},
\end{equation}
\begin{equation}
S^V=-\frac 1{6\sqrt{15}}A(\overline{k})\overline{k}^2\left( 0.1330+0.00171%
\overline{k}^2\right) ,
\end{equation}
\begin{equation}
L^V=\sqrt{3}A(\overline{k})\frac km\left( -0.00171+0.000549\overline{k}%
^2\right) ,
\end{equation}
\begin{equation}
\left( M1\right) ^A=\frac 1{\sqrt{3}}A(\overline{k})\left( 2.1114+0.0567%
\overline{k}^2-0.00084\overline{k}^4\right) ,
\end{equation}
\begin{equation}
\left( E2\right) ^A=\frac 1{\sqrt{3}}A(\overline{k})\left(
0.0521\overline{k}%
^2-0.236\overline{k}^4\right) \times 10^{-3},
\end{equation}
\begin{equation}
L^A=-A(\overline{k})\left( 0.862+0.0189\overline{k}^2-0.000737\overline{k}%
^4\right) ,
\end{equation}
\begin{equation}
S^A=\sqrt{\frac 23}\frac kmA(\overline{k})\left( 0.337+0.00480\overline{k}%
^2-0.000172\overline{k}^4\right) .
\end{equation}
This gives us a feeling of the sensitivity of the weak helicity amplitudes
to small changes in the IK wave functions. The form-factors calculated from
this choice is labeled as ``Isgur-Karl 2'' in Figs. 1-11.

We also include the result of a D-state mixing model, suggested by Glashow
\cite{46}, and further discussed by VBJ \cite{46}, Bourdeau and Mukhopadhyay
$\cite{32}:$%
\begin{equation}
|N>=\sqrt{1-\gamma }|^2S_s>+\sqrt{\gamma }|^4D_M>,
\end{equation}
\begin{equation}
|\Delta >=\sqrt{1-3\beta }|^4S_s>-\sqrt{2\beta }|^4D_s>+\sqrt{\beta }%
|^2D_M>.
\end{equation}
The purpose here is to get the correct $g_A$ by changing $\gamma $, and
adjust $\beta $ so that the $EMR$ comes out to be the PDG recommended value.
We find
\begin{equation}
\gamma =0.2048\pm 0.0015,\;\beta =0.103\pm 0.011,
\end{equation}
and together they give:
\begin{equation}
g_A=1.257\pm 0.003,\;EMR=-0.015\pm 0.004.
\end{equation}
The results of this model is included in Table \ref{key4} and Figs. 1-11.

A recent experiment \cite{37} shows that $EMR=-0.026$. This implies, in the
D-state mixing model, $\beta =0.073$. For brevity, we do not give numerical
results for the helicity amplitudes for this case.

\subsection{The $\left( E2\right) ^A/\left( M1\right) ^A$ ratio}

The ratio $E2/M1$ in the $\gamma N\rightarrow \Delta $ transition has
attracted a lot of attention of theorists \cite{25,36,37} and
experimentalists \cite{38} alike. In analogy to this quantity, we can also
define the ratio
\begin{equation}
AEMR\equiv \left( E2\right) ^A/\left( M1\right) ^A,
\end{equation}
for the axial-vector analogue of the EMR for the process (3). In the SU(6)
symmetry limit, it is zero identically.

We give in Fig. 11 a plot of the ratio as a function of $Q^2$, keeping W
fixed at the Delta excitation, as computed from the QM, and compare it with
the Adler assumption that it is zero identically, an assumption made in the
Kitagaki {\it et al. }\cite{18} analysis. It would be interesting to
determine this quantity directly from the experiment in the future. We
recall here that the $EMR$ , for the vector current, is not well-determined
in the QM of the IK variety. While Davidson {\it et al}. \cite{25} found the
value for this ratio at the photon point to be
\begin{equation}
EMR\left( Q^2=0,\cite{25}\right) =-0.0157\pm 0.0072,
\end{equation}
from the pion photoproduction in the Delta region. New experiments \cite{37}
at the Brookhaven LEGS facility have indicated this ratio to be at least
{\em twice} as large. The IK quark model predicts the ratio at the photon
point
\begin{equation}
EMR\left( Q^2=0;IK\;Model\right) =-0.0033.
\end{equation}
The alternative wave-functions of IK model (Isgur-Karl 2), given above
(Eq.(105-106)), yield an EMR of -0.0051. Thus, the EMR, obtained in the IK
quark model, is much smaller in magnitude compared to the phenomenological
value. There are indications in the latest Skyrmion approach \cite{13} that
the EMR at the photon point is indeed much larger than the prediction of the
IK quark model, close to -0.05. Thus, the experimental determination of the
AEMR may be less difficult than what the quark model suggests, as the latter
may turn out to be a gross underestimate. This quantity should be computed
in the soliton models.

While we are on the subject of the EMR and AEMR, we should make some
remarks about the issue of the behavior of these ratios as functions of
$Q^2$, a
very topical question. In the perturbative QCD approach (pQCD) \cite{40},
these ratios must approach unity, as $Q^2\rightarrow \infty $:
\begin{equation}
EMR\left( Q^2\rightarrow \infty \right) \Rightarrow 1,\;AEMR\left(
Q^2\rightarrow \infty \right) \Rightarrow 1,
\end{equation}
while in the IK model, these behave as follows:
\begin{equation}
EMR=-\frac{6.26+0.627\overline{k}^2+0.252\overline{k}^4}{2.0237+0.0623%
\overline{k}^2-0.00133\overline{k}^4}\times 10^{-3},
\end{equation}
\begin{equation}
AEMR=\frac{0.325\overline{k}^2-0.251\overline{k}^4}{2.0204+0.0625\overline{k}%
^2-0.00133\overline{k}^4}\times 10^{-3}.
\end{equation}
The last two equations are, of course, meaningless at large $Q^2$ (that is,
large $\overline{k}^2$). We are giving the last two equations merely to
indicate very different limits for the EMR and AEMR, at large $Q^2,$ in the
IK quark model, compared with pQCD. Nevertheless, it is amusing to note that
the IK quark model reproduces the signs and rough orders of magnitudes of
the pQCD limits mentioned above.

Several questions are important here: (1) Do the EMR and AEMR approach unity
for large $Q^2$ attainable in the laboratory? If the answer is yes, at what
$%
Q^2$? (2) Where does the IK-type quark model predictions definitely break
down? (3) What happens to the Bloom-Gilman duality \cite{41} in the axial
vector sector, on which we have no experimental information? Future research
should focus on these issues. We may add here that there is a strong debate
in the literature on the question (1). The pQCD ``believers'' \cite{40} tend
to see the asymptopia at hand at $Q^2\geq 6\,GeV^2$, while the
``non-believers'' \cite{42} argue that the $Q^2$ value for the pQCD rules to
be valid is too large for us to worry, at the ``modest'' Q$^2$ currently
available.

\subsection{Further comments on the comparison of our work with others}

Previous to our work, many authors have investigated the neutrino excitation
of the Delta resonance off nucleons, but all in the context of the SU(6)
symmetric quark model \cite{5,6,7,8}. Schreiner and von Hippel \cite{10}
have reviewed most of these attempts. Among the later works, mention may be
made of the work of Abdullah and Close \cite{8}. In this work, the CVC
condition is imposed, also, quark structure functions are introduced, and
the quark form factors are treated as constants. In the work by Andreadis
{\it et al}. \cite{11}, again in the SU(6) limit, quark form factors are
introduced. We do not use form factors at the quark level for the following
reason: (1) we want to test quark model predictions rather than fitting a
constituent-quark form-factor. (2) We believe that the effects of the gluons
and sea-quarks have already been absorbed in the effective potential of
Isgur-Karl quark model. (3) Our results {\em without} the form-factor is
very good compared to the currently available experiments. Thus, there is no
phenomenological reason to introduce quark form factors. (4) Introduction of
quark-level form factors would complicate even more the relationship of the
quark model to QCD. As it is, this relationship is far from clear.

Finally, we should discuss the difference between our approach here and a
recent work by Hemmert, Holstein and Mukhopadhyay(HHM) \cite{14} dealing
with the NN and N$\Delta $ couplings in the quark model (see Table
\ref{key4}%
). In that work, relativistic corrections to the nucleon $g_A$ are taken
into account and an agreement with the experimental $g_A$ for the nucleon is
reached. But it yields an off-diagonal N$\Delta $ axial coupling
substantially lower than the experimental values. In contrast to HHM, the
non-relativistic IK approach, used here, does not have the relativistic
correction taken into account. Thus, the diagonal value of $g_A$ still
remains off the experimental value ($g_A(IK)=1.63$). But the off-diagonal
axial vector matrix elements come out better. Overall, {\em the color
hyperfine interaction does not remove the discrepancy between quark model
estimates and experiments in either approach, when both diagonal and
off-diagonal effects are computed}. Thus, the problem of the quark model in
simultaneously explaining diagonal {\em and} off-diagonal observables does
not disappear.

\section{Summary and conclusions}

We have computed weak amplitudes of the $N\rightarrow \Delta $ transition in
the framework of the IK quark model, thereby dealing with the effect of the
color-hyperfine interactions in these amplitudes. Our main conclusions are:

\begin{enumerate}
\item  The deficit of the nucleon to Delta (1232) magnetic dipole amplitude
in the IK quark model estimate, when compared with pion photoproduction
analysis \cite{25}, is confirmed via the vector form factor $C_3^V$ at $%
q^2\rightarrow 0$. This deficit seems to heal around of $Q^2=0$.25 GeV$^2$.

\item  There is a mild violation of the CVC in the IK model. Thus, the
amplitude $C_6^V$ , which should be zero by CVC, is predicted in the IK
quark model to have a small but non-zero value.

\item  The axial-vector transverse amplitudes are largely well-described in
the IK model. An exception is our inability to get the PCAC value of the $%
C_6^A.$ This is not surprising, since we do not have explicit meson degrees
of freedom. The IK model also violates axial current conservation in the
chiral limit. Despite this shortcoming, we are able to reproduce the
off-diagonal Goldberger-Treiman estimate of the $C_5^A(0)$ in the IK quark
model.

\item  There is a simple way of parameterizing the SU(6) breaking effects,
through a relation (Eq.(56)) that connects the two transverse helicity
amplitudes to the Adler form factor $C_3^A$. The IK model gives an estimate
of this small effect. Its experimental verification, though indirect and
difficult, should constitute an important experimental challenge, analogous
to the determination of E2/M1 in the vector (electromagnetic) sector.
\end{enumerate}

Hopefully, new weak interactions studies in the nucleon resonance region
(1-2 GeV of W) will be possible at existing neutrino facilities. There are
also new experimental possibilities at the CEBAF on the weak charged and
neutral current explorations [43-47] of the isobar physics. Though these are
intrinsically very difficult, there is some hope that such experiments would
be possible. They would go a long way towards our understanding of the axial
vector response of the nucleon, in particular. The role of the Delta isobar,
already important in the Adler-Weissberger sum-rule \cite{48}, would be
interesting to be explored further in the nucleon to Delta weak excitation
domain.

\vspace{0.2in}

{\Large {\bf Acknowledgments}}

One of us (NCM) thanks T. Hemmert, B. Holstein and J. Napolitano for many
helpful discussions. This research has been supported by the U.S. Dept. of
Energy.

\newpage
\textwidth 6.3in
\begin{table}
\begin{tabular}{|c|cc|} \hline
 &
$\sqrt{\frac 32}L_{1/2}^A$ & $\sqrt{\frac 32}\frac{3m}kS_{1/2}^A$ \\ \hline
$<\Delta ^4S_s|N^2S_s>$ & $-\frac 2{\sqrt{3}}$ & $\frac 1{3 \sqrt{3}}$ \\
$<\Delta ^4S_s|N^2S_{s^{\prime }}>$ & $\frac 19\overline{k}^2$ & $
-\frac 29\left( 1+\frac 1{12}
\overline{k}^2\right)$ \\ $<\Delta ^4S_{s^{\prime }}|N^2S_s>$ & $\frac 19%
\overline{k}^2$ & $\frac 29\left( 1-\frac 1{12}
\overline{k}^2\right)$ \\ $<\Delta ^4S_s|N^2S_M>$ & $\frac
1{9\sqrt{2}}\overline{k}^2$ & $-
\frac{\sqrt{2}}9\left( 1+\frac 1{12}\overline{k}^2\right)$ \\ $<\Delta
^4S_s|N^4D_M>$ & $\frac 1{18\sqrt{5}}\overline{k}^2$ & $-\frac 1{6
\sqrt{5}}-\frac 1{9\sqrt{5}}\left( 1+\frac 1{12}\overline{k}^2\right)$ \\
$<\Delta ^4D_s|N^2S_s>$ & $-\frac 19\sqrt{\frac 25}\overline{k}^2$ & $-\frac
13
\sqrt{\frac 25}-\frac 29\sqrt{\frac 25}\left( 1-\frac 1{12}\overline{k}%
^2\right)$ \\ $<\Delta ^2D_M|N^2S_s>$ & $\frac 1{18\sqrt{5}}\overline{k}^2$ &
$
\frac 1{6
\sqrt{5}}+\frac 1{9\sqrt{5}}\left( 1-\frac 1{12}\overline{k}^2\right)$ \\
$<\Delta ^4S_{s^{\prime }}|N^2S_{s^{\prime }}>$ & $\frac 1{6\sqrt{3}}\left(
-12+\frac 43\overline{k}^2-\frac 19\overline{k}^4\right)$ & $\frac 1{3
\sqrt{3}}\left( 1-\frac 19\overline{k}^2+\frac 1{108}\overline{k}^4\right)$
\\ $<\Delta ^4S_{s^{\prime }}|N^2S_M>$ & $\frac 1{6\sqrt{6}}\left( \frac 43%
\overline{k}^2-\frac 19\overline{k}^4\right)$ &
$-\frac 1{27\sqrt{6}}\left(\overline{k}^2-\frac 1{12}\overline{k}^4\right)$
\\ $<\Delta
^4S_{s^{\prime }}|N^4D_M>$ & $\frac 1{9\sqrt{15}}\left( \overline{k}^2-\frac
1{12}\overline{k}^4\right)$ & $\frac 1{36
\sqrt{15}}\overline{k}^2-\frac 1{54\sqrt{15}}\left( \overline{k}^2-\frac
1{12}\overline{k}^4\right)$ \\ $<\Delta ^4D_s|N^2S_{s^{\prime }}>$ & $-\frac
29%
\sqrt{\frac 2{15}}\left( \overline{k}^2-\frac 1{12}\overline{k}^4\right)$ &
$\frac 1{18}
\sqrt{\frac 2{15}}\overline{k}^2+\frac 1{27}\sqrt{\frac 2{15}}\left(
\overline{k}^2-\frac 1{12}\overline{k}^4\right)$ \\ $<\Delta ^4D_s|N^2S_M>$ &
$-\frac 2{9\sqrt{15}}\left( \overline{k}^2-\frac 1{12}\overline{k}^4\right)$
&
$\frac 1{18
\sqrt{15}}\overline{k}^2+\frac 1{27\sqrt{15}}\left( \overline{k}^2-\frac
1{12}\overline{k}^4\right)$ \\ $<\Delta ^4D_s|N^4D_M>$ &
$-\frac 1{5\times 18\sqrt{6}}\left( 13\overline{k}^2-\frac
13\overline{k}^4\right)$ &
$\frac{\sqrt{6}}{12}\left( 1-\frac 1{15}\overline{k}^2\right) +\frac
1{540\sqrt{6}}
\left( 13\overline{k}^2-\frac 13\overline{k}^4\right)$ \\
$<\Delta^2D_M|N^2S_{s^{\prime }}>$ & $\frac 1{9\sqrt{15}}\left(
\overline{k}^2-\frac
1{12}\overline{k}^4\right)$ & $-\frac 1{36
\sqrt{15}}\overline{k}^2-\frac 1{54\sqrt{15}}\left( \overline{k}^2-\frac
1{12}\overline{k}^4\right)$ \\ $<\Delta ^2D_M|N^2S_M>$ &
$\frac 1{12\sqrt{30}}\left( \frac{16}3\overline{k}^2-\frac
19\overline{k}^4\right)$ &
$-\frac 1{36\sqrt{30}}\overline{k}^2-\frac 1{54\sqrt{30}}\left(
4\overline{k}^2-\frac
1{12}\overline{k}^4\right)$ \\ $<\Delta ^2D_M|N^4D_M>$ &
$\frac 1{30\sqrt{3}}\left( 30-\frac{17}3\overline{k}^2+\frac
29\overline{k}^4\right)$ &
$\frac 1{2\sqrt{3}}\left( 1-\frac 1{15}\overline{k}^2\right) -\frac
1{54\sqrt{3}}
\left( 9-\frac{17}{10}\overline{k}^2+\frac 1{15}\overline{k}%
^4\right)$ \\ \hline
\end{tabular}
\caption{Axial-vector longitudinal and scalar amplitudes for various nucleon
to Delta SU(6)
configurations. Common factor is
$\protect\sqrt{\frac{4\pi \alpha
_W}{2K_0}}e^{-\overline{k}^2/6}.$\label{key1}}
\end{table}

\begin{table}
\begin{tabular}{|c|cc|} \hline
 &
$\sqrt{3}A_{1/2}^A$ & $A_{3/2}^A$ \\  \hline
$<\Delta ^4S_s|N^2S_s>$ & $-\frac 2{\sqrt{3}}$ & $-\frac 2{\sqrt{3}}$ \\
$<\Delta ^4S_s|N^2S_{s^{\prime }}>$ & $\frac 19\overline{k}^2$ & $
\frac 19  \overline{k}^2$ \\
$<\Delta ^4S_{s^{\prime }}|N^2S_s>$ & $\frac 19\overline{k}^2$
& $\frac 19\overline{k}^2$ \\
$<\Delta ^4S_s|N^2S_M>$ & $\frac 1{9\sqrt{2}}\overline{k}^2$ & $
\frac 1{9\sqrt{2}}\overline{k}^2$ \\
$<\Delta ^4S_s|N^4D_M>$ & $\frac 2{9\sqrt{5}}$ & $-\frac 1{9
\sqrt{5}}\overline{k}^2$ \\
$<\Delta ^4D_s|N^2S_s>$ & $-\frac 2{9\sqrt{10}}%
\overline{k}^2$ & $\frac 2{9\sqrt{10}}\overline{k}^2$ \\
$<\Delta ^2D_M|N^2S_s>$ & $-\frac 1{9\sqrt{5}}%
\overline{k}^2$ & $0$ \\
$<\Delta ^4S_{s^{\prime }}|N^2S_{s^{\prime }}>$ & $\frac 1{6\sqrt{3}}\left(
-12+\frac 43\overline{k}^2-\frac 19\overline{k}^4\right)$ & $\frac 1{6
\sqrt{3}}\left( -12+\frac 43\overline{k}^2-\frac 19\overline{k}^4\right)$ \\
$<\Delta ^4S_{s^{\prime }}|N^2S_M>$ & $\frac 1{6\sqrt{6}}\left( \frac 43%
\overline{k}^2-\frac 19\overline{k}^4\right)$ & $\frac 1{6
\sqrt{6}}\left( \frac 43\overline{k}^2-\frac 19\overline{k}^4\right)$ \\
$<\Delta ^4S_{s^{\prime }}|N^4D_M>$ & $\frac 1{3\sqrt{15}}\left( \frac 43%
\overline{k}^2-\frac 19\overline{k}^4\right)$ & $\frac 1{6
\sqrt{15}}\left( \frac 43\overline{k}^2-\frac 19\overline{k}^4\right)$ \\
$<\Delta ^4D_s|N^2S_{s^{\prime }}>$ & $-\frac 1{3\sqrt{30}}\left( \frac 43%
\overline{k}^2-\frac 19\overline{k}^4\right)$ & $\frac 1{3
\sqrt{30}}\left( \frac 43\overline{k}^2-\frac 19\overline{k}^4\right)$ \\
$<\Delta ^4D_s|N^2S_M>$ & $-\frac 1{6\sqrt{15}}\left( \frac 43\overline{k}%
^2-\frac 19\overline{k}^4\right)$ & $\frac 1{6
\sqrt{15}}\left( \frac 43\overline{k}^2-\frac 19\overline{k}^4\right)$ \\
$<\Delta ^4D_s|N^4D_M>$ & $-\frac 1{15\sqrt{6}}\left( \frac{17}3\overline{k}%
^2-\frac 29\overline{k}^4\right)$ & $-\frac 1{15
\sqrt{6}}\left( \frac{10}3\overline{k}^2-\frac 19\overline{k}^4\right)$ \\
$<\Delta ^2D_M|N^2S_{s^{\prime }}>$ & $-\frac 1{6\sqrt{15}}\left( \frac 43%
\overline{k}^2-\frac 19\overline{k}^4\right)$ & $0$ \\
$<\Delta ^2D_M|N^2S_M>$ & $-\frac 1{6\sqrt{30}}\left( \frac{16}3\overline{k}%
^2-\frac 19\overline{k}^4\right)$ & $0$ \\
$<\Delta ^2D_M|N^4D_M>$ & $\frac 1{30\sqrt{3}}\left(
30-\frac{17}3\overline{k}%
^2+\frac 29\overline{k}^4\right)$ & $\frac 1{\sqrt{3}}\left( 1-\frac 1{30}%
\overline{k}^2\right)$ \\ \hline
\end{tabular}
\caption{Transverse axial-vector amplitudes for various nucleon to Delta
SU(6)
configurations. Common factor is
$\protect\sqrt{\frac{4\pi \alpha
_W}{2K_0}}e^{-\overline{k}^2/6}$.\label{key2}}
\end{table}

\begin{table}
\begin{tabular}{|c|cccc|} \hline
 & $\frac
mkL_{1/2}^V$ & $S_{1/2}^V$ & $\sqrt{3}A_{1/2}^V(H_{II}^V)$ &
$A_{3/2}^V(H_{II}^V)$
\\  \hline
$<\Delta ^2D_M|N^2S_s>$ & $-\frac 1{3\sqrt{15}}\left(
1-\frac{\overline{k}^2}{12}\right) $ &
$\frac 1{6\sqrt{15}}\overline{k}^2$ & $\frac 1{\sqrt{5}}$ & $-\frac
1{3\sqrt{5}}$ \\ $<\Delta ^2D_M|N^2S_{s^{\prime }}>$ & $\frac
1{54\sqrt{5}}\left(
\overline{k}^2-\frac{\overline{k}^4}{12}\right) $ & $\frac 1{9\sqrt{5}}\left(
\overline{k}^2-\frac{\overline{k}^4}{12}\right) $ & $-\frac 1{6\sqrt{15}}%
\overline{k}^2$ & $\frac 1{18
\sqrt{15}}\overline{k}^2$ \\ $<\Delta ^2D_M|N^2S_M>$ & $-\frac
1{648\sqrt{10}}%
\overline{k}^4$ & $-\frac 1{108\sqrt{10}}\overline{k}^4$ & $-\frac
1{6\sqrt{30}}%
\overline{k}^2$ & $\frac 1{18
\sqrt{30}}\overline{k}^2$ \\ $<\Delta ^4S_s|N^4D_M>$ &
$-\frac 1{3\sqrt{15}}\left( 1+\frac{\overline{k}^2}{12}\right) $ &
$-\frac 1{6\sqrt{15}}\overline{k}^2$ & $\frac 1{\sqrt{5}}$ & $-\frac 1{3
\sqrt{5}}$ \\ $<\Delta ^4S_{s^{\prime }}|N^4D_M>$ & $-\frac
1{54\sqrt{5}}\left(
\overline{k}^2-\frac{\overline{k}^4}{12}\right) $ & $-\frac
1{9\sqrt{5}}\left(
\overline{k}^2-\frac{\overline{k}^4}{12}\right) $ & $-\frac 1{6\sqrt{15}}%
\overline{k}^2$ & $\frac 1{18
\sqrt{15}}\overline{k}^2$ \\ $<\Delta ^4D_s|N^4D_M>$ &
$\frac{\sqrt{2}}{1080}%
\left( 7\overline{k}^2-\frac{\overline{k}^4}3\right) $ &
$\frac{\sqrt{2}}{180}%
\left( 7\overline{k}^2-\frac{\overline{k}^4}3\right) $ &
$-\frac 1{\sqrt{6}}\left( 1-\frac{\overline{k}^2}{15}\right) $ & $-\frac
1{\sqrt{6}}\left( 1-%
\frac{\overline{k}^2}{15}\right) $ \\ \hline
\end{tabular}
\caption{Vector amplitudes amplitudes for various nucleon to Delta SU(6)
configurations. Common factor is
$\protect\sqrt{\frac{4\pi \alpha _W}{2K_0}}e^{-\overline{k}^2/6}$.
 The transverse contribution listed is from
$H_2^V$ only. Transverse contribution from $H_1^V$ equals to its axial
counterpart (Table 2) times $k/2m$. Contributions
from the other configurations not shown are zero. \label{key3}}
\end{table}

\begin{table}
\begin{tabular}{|c|cccc|cccc|} \hline
 & $C_3^V$ & $C_4^V$ & $C_5^V$ & $C_6^V$ & $C_3^A$ & $C_4^A$ & $C_5^A$ & $
C_{6,non-pole}^A$ \\ \hline
Salin \cite{3} & 2.0 & 0 & 0 & 0 & 0 & -2.7 & 0 & 0 \\
Adler \cite{4} & 1.85 & -0.89 & 0 & 0 & 0 & -0.3 & 1.2 & 0 \\
Bijtebier \cite{5} & 2.0 & 0 & 0 & 0 & 0 & -2.9$\sim$ -3.6 & 1.2 & 0 \\
Zucker \cite{7} & - & - & - & 0 & 1.8 & 1.8 & 1.9 & 0 \\
HHM \cite{14} & 1.39 & - & - & 0 & 0 & -0.29$\pm$ 0.006 & 0.87$\pm$ 0.03 & -
\\
SU(6) & 1.48 & -1.13 & 0 & 0 & 0 & -0.38 & 1.17 & 0 \\
Isgur-Karl & 1.32 & -0.79 & -0.36 & 0.014 & -0.0013 & -0.66 & 1.16 &
0.032 \\
Isgur-Karl 2 & 1.37 & -0.66 & -0.59 & -0.015 & 0.0008 & -0.657 & 1.20
& 0.042 \\
D-mixing \cite{32,36} & 1.29 & 0.78 & -1.9 & -0.15 & 0.052 & 0.052 & 0.813 &
-0.17\\
\hline
\end{tabular}
\caption{A comparison of the Adler form factors in the $N\rightarrow \Delta
$ weak transition in different approaches indicated in the text. The momentum
transfer squared is taken to be zero. The last four rows are from this
work.\label{key4}}
\end{table}
\newpage\ \vspace{0.4in}

\noindent {\Large {\bf The Figure Captions}}

\begin{description}
\item[Figure 1:]  The ratio $A_{3/2}^V/\sqrt{3}A_{1/2}^V$ vs. $Q^2$.

\item[Figure 2:]  The ratio $A_{3/2}^A/\sqrt{3}A_{1/2}^A$ vs. $Q^2$.

\item[Figure 3:]  $C_3^V$ as a function of $Q^2$. Results from the two
versions of the IK quark-model, SU(6) limit and the D-wave mixing model, are
compared with the experimetal results of Kitagaki {\it et al}\cite{18}. The
experimental errors shown here are due to the uncertainty in $M_V$
(Eq.(93))$%
.$

\item[Figure 4:]  $C_4^V$ as a function of $Q^2$.

\item[Figure 5:]  $C_5^V$ vs. $Q^2$. This form-factor is assumed to be zero
in experimental fits. Its non-zero value here indicates the violation of the
{\it magnetic dipole dominance.}

\item[Figure 6:]  $C_6^V$ vs. $Q^2$. Its non-zero value indicates the degree
of CVC violation in the quark model.

\item[Figure 7:]  $C_3^A$ vs. $Q^2$. Its non-zero value indicates the
violation of {\it magnetic dipole dominance.}

\item[Figure 8:]  $C_4^A$ vs. $Q^2$.

\item[Figure 9:]  $C_5^A$ vs. $Q^2$. Results from two versions of the IK
quark-model, SU(6) limit and D-wave mixing model are compared with
experimental results of Kitagaki {\it et al}. The experimental errors shown
are generated by the uncertainty in the fitted parameter $M_A$ (Eq.(98)).

\item[Figure 10:]  $C_6^A$ vs. $Q^2$. Since the quark model does not have
pion-pole term built in it, this is only the non-pole contribution.

\item[Figure 11:]  $E2/M1$ and $(E2)^A/(M1)^A$, EMR and AEMR respectively,
vs. $Q^2$.
\end{description}

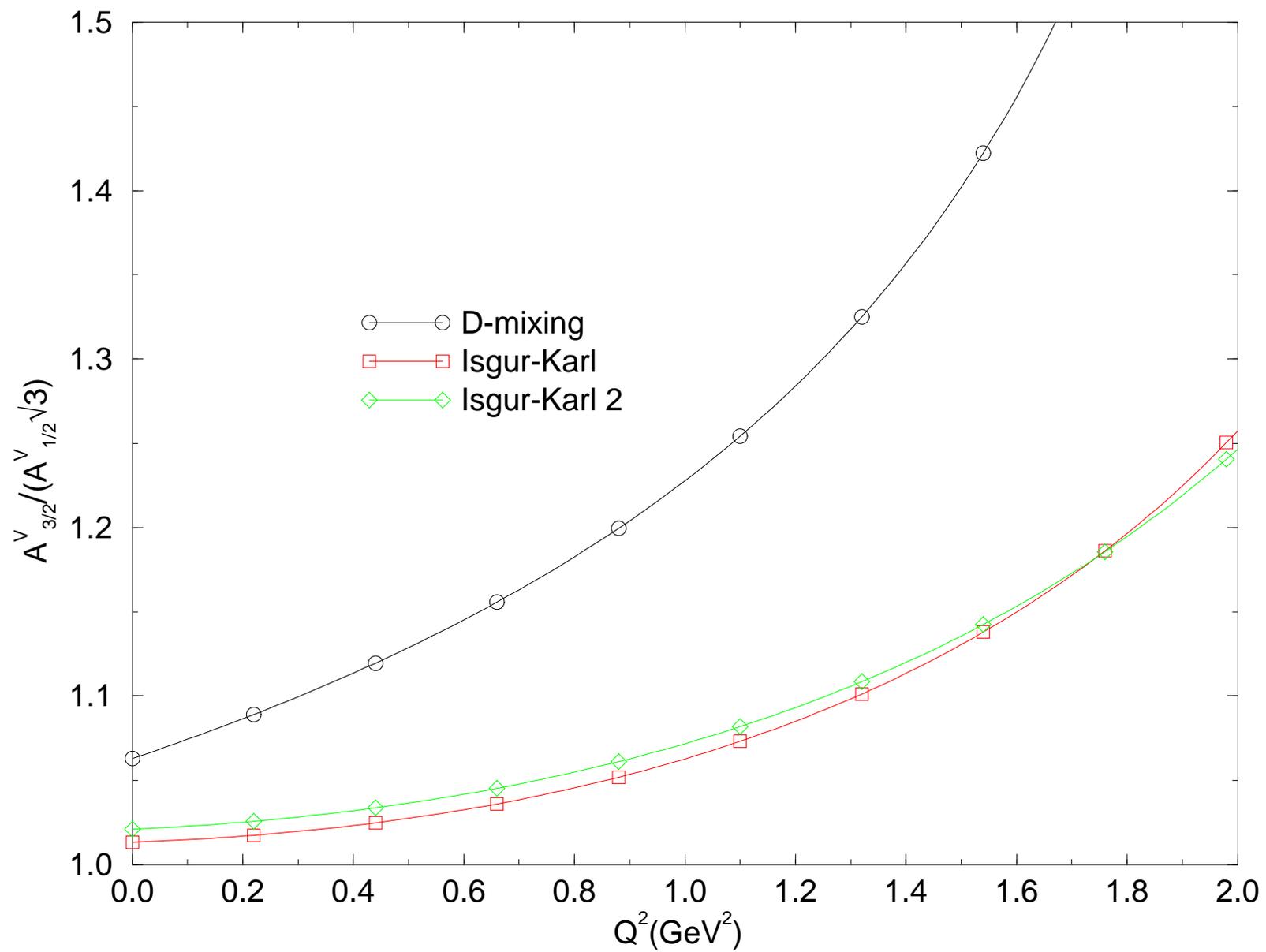
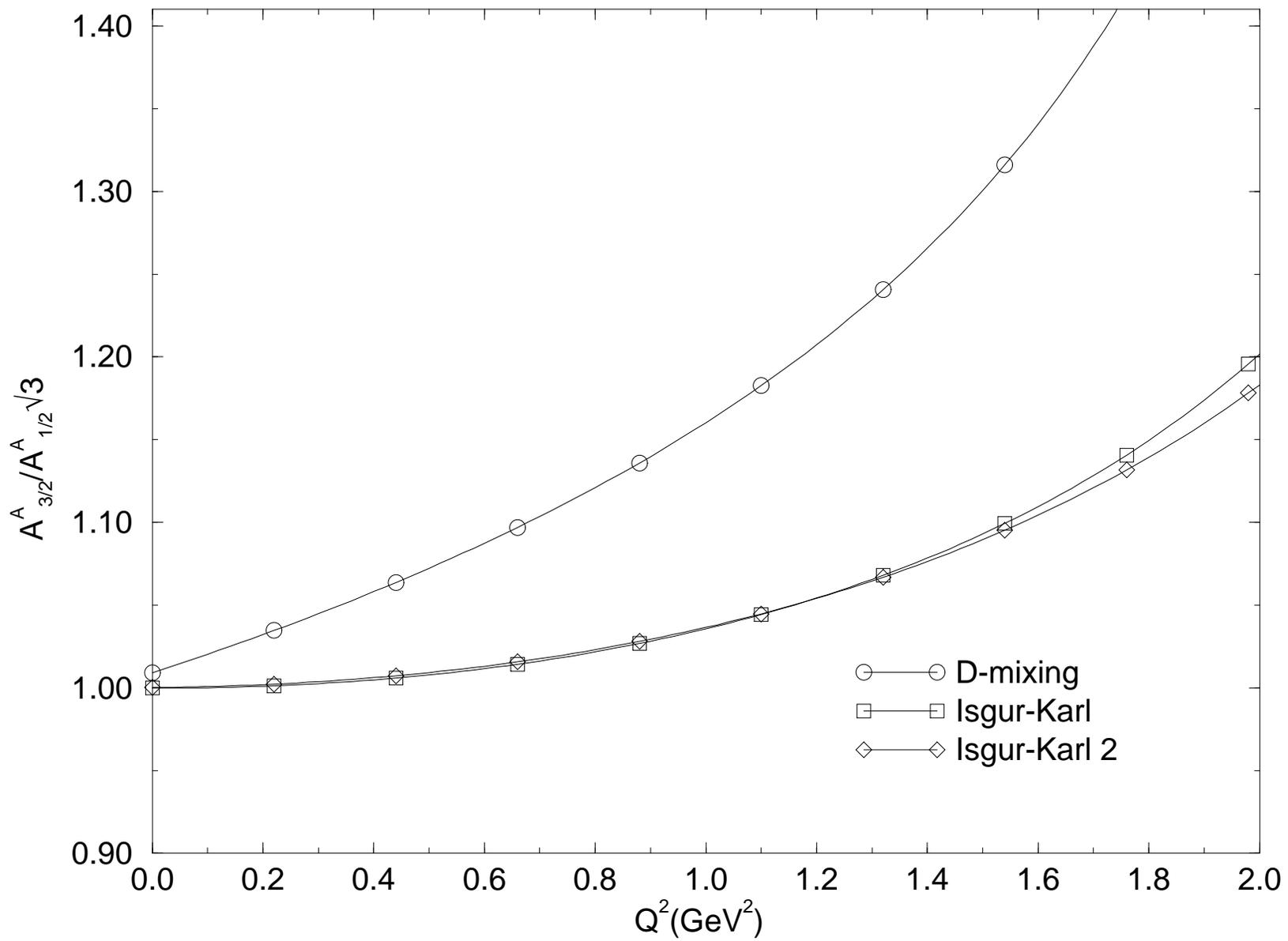
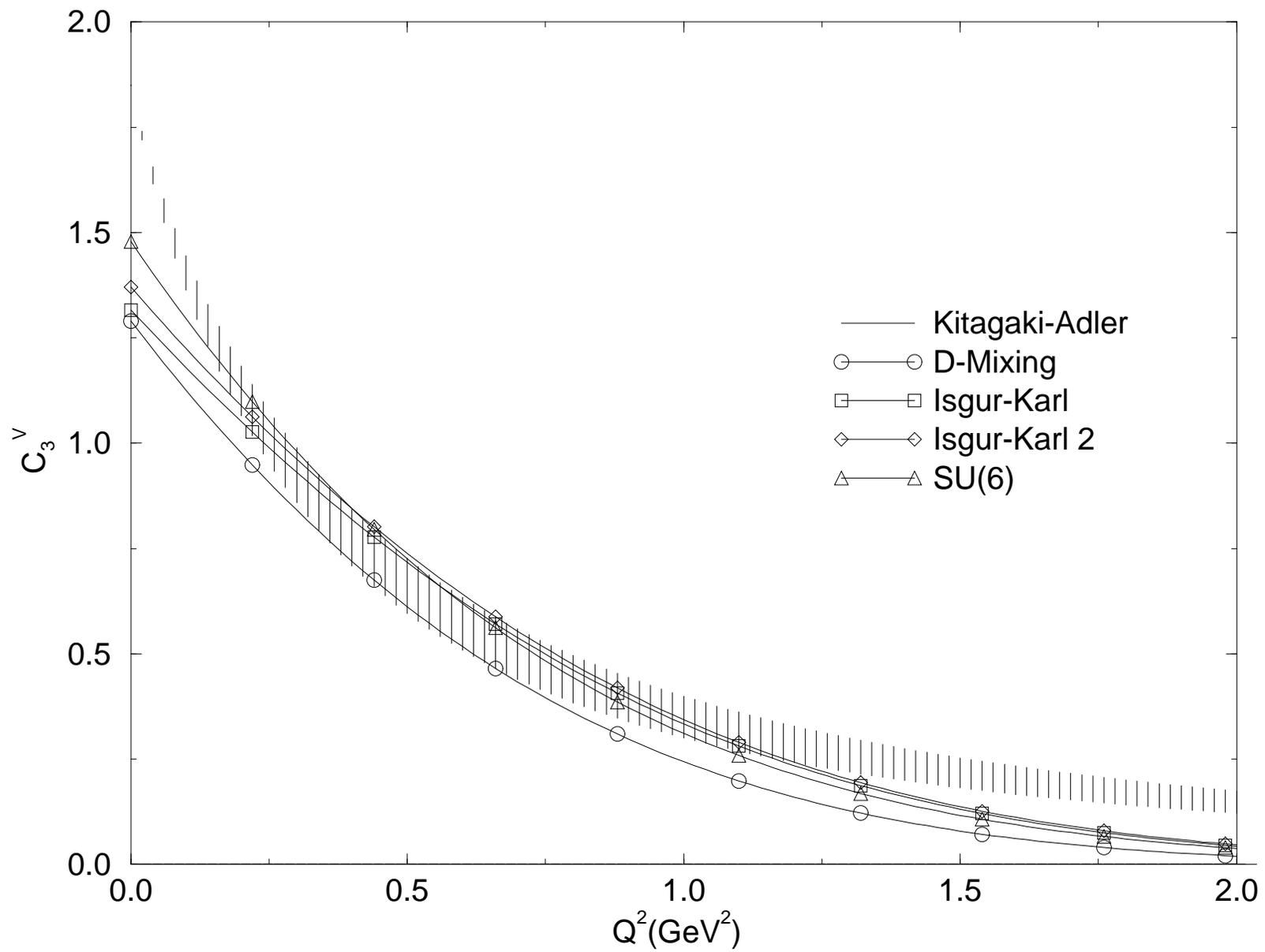
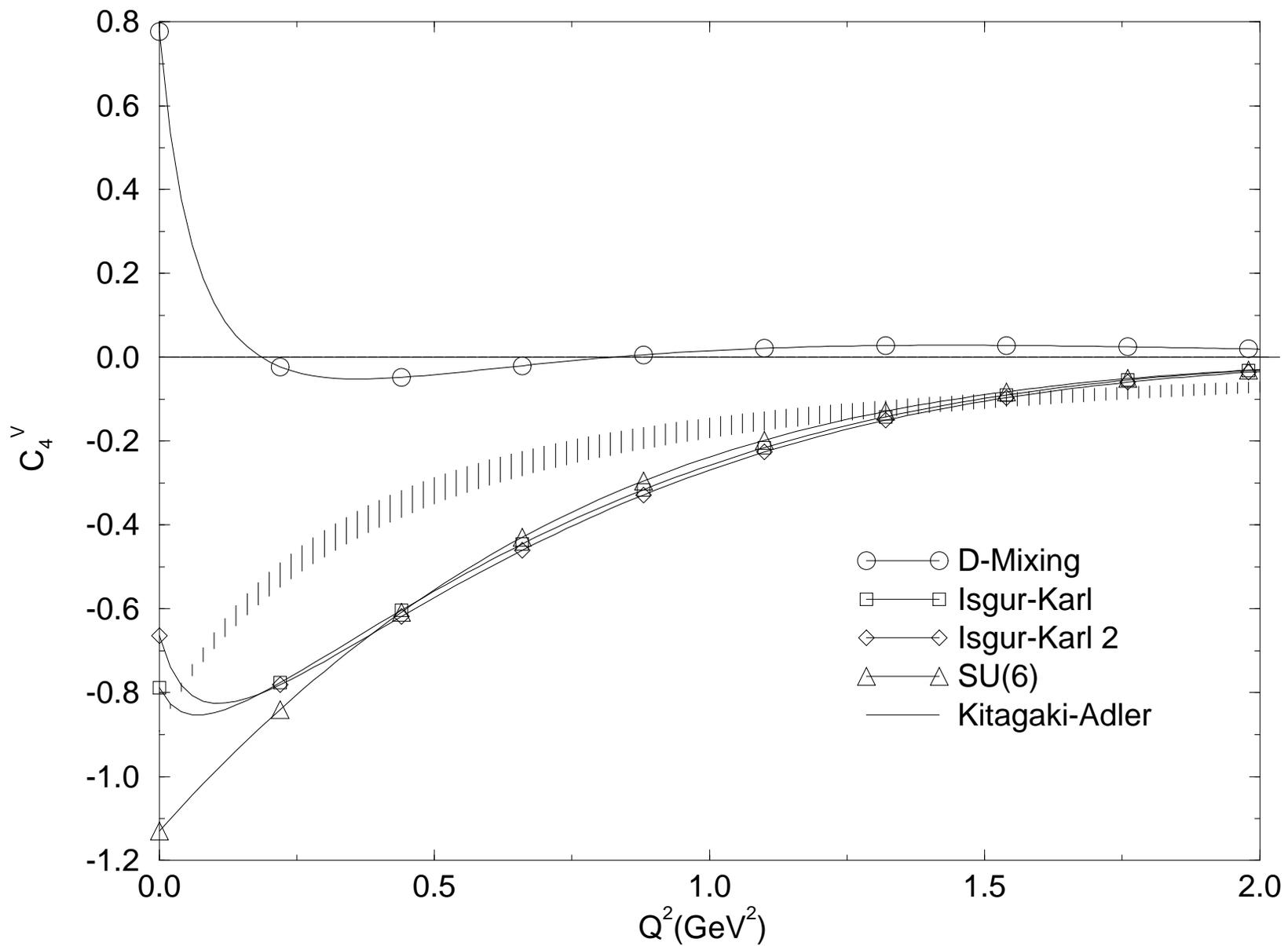
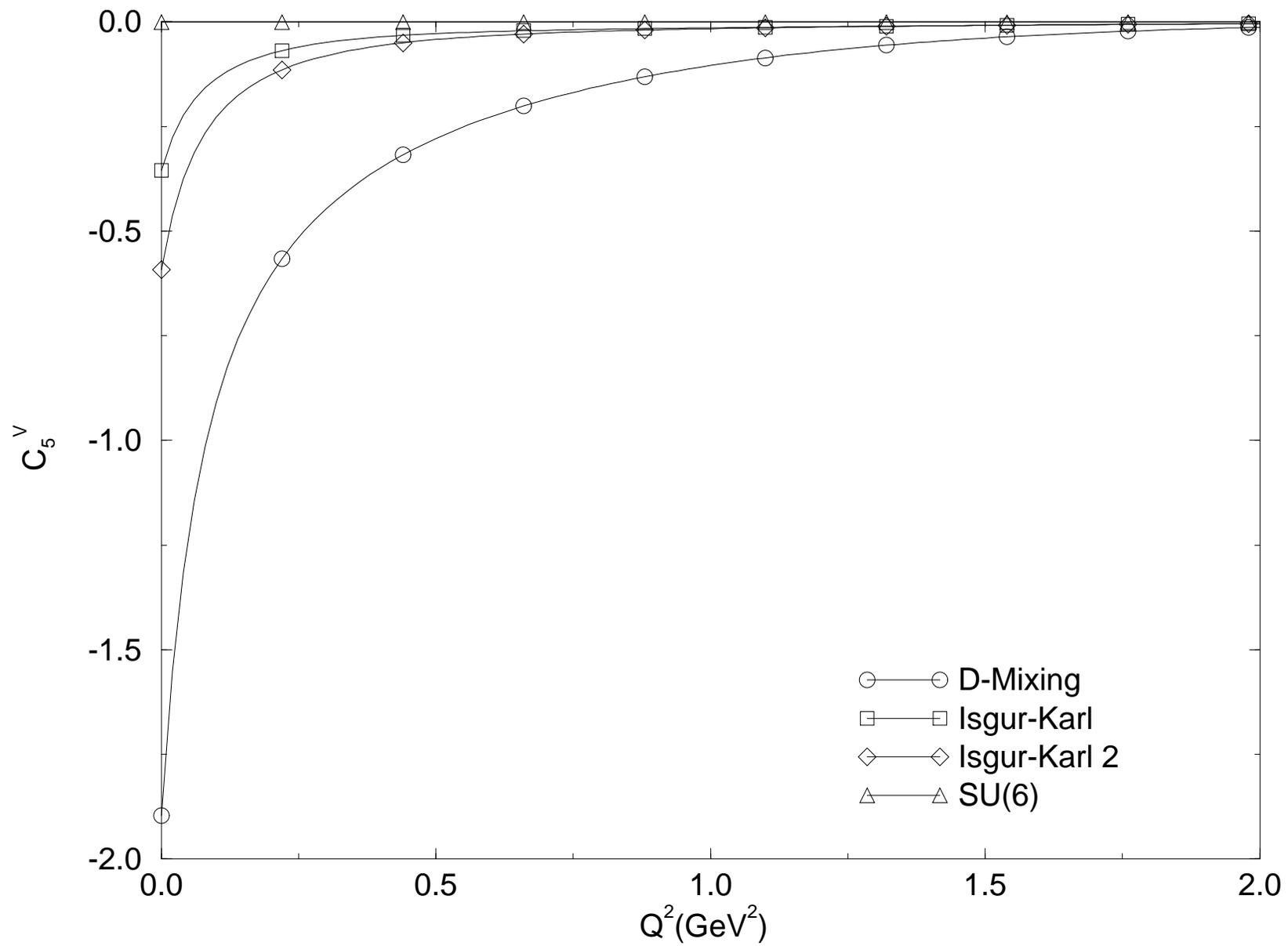
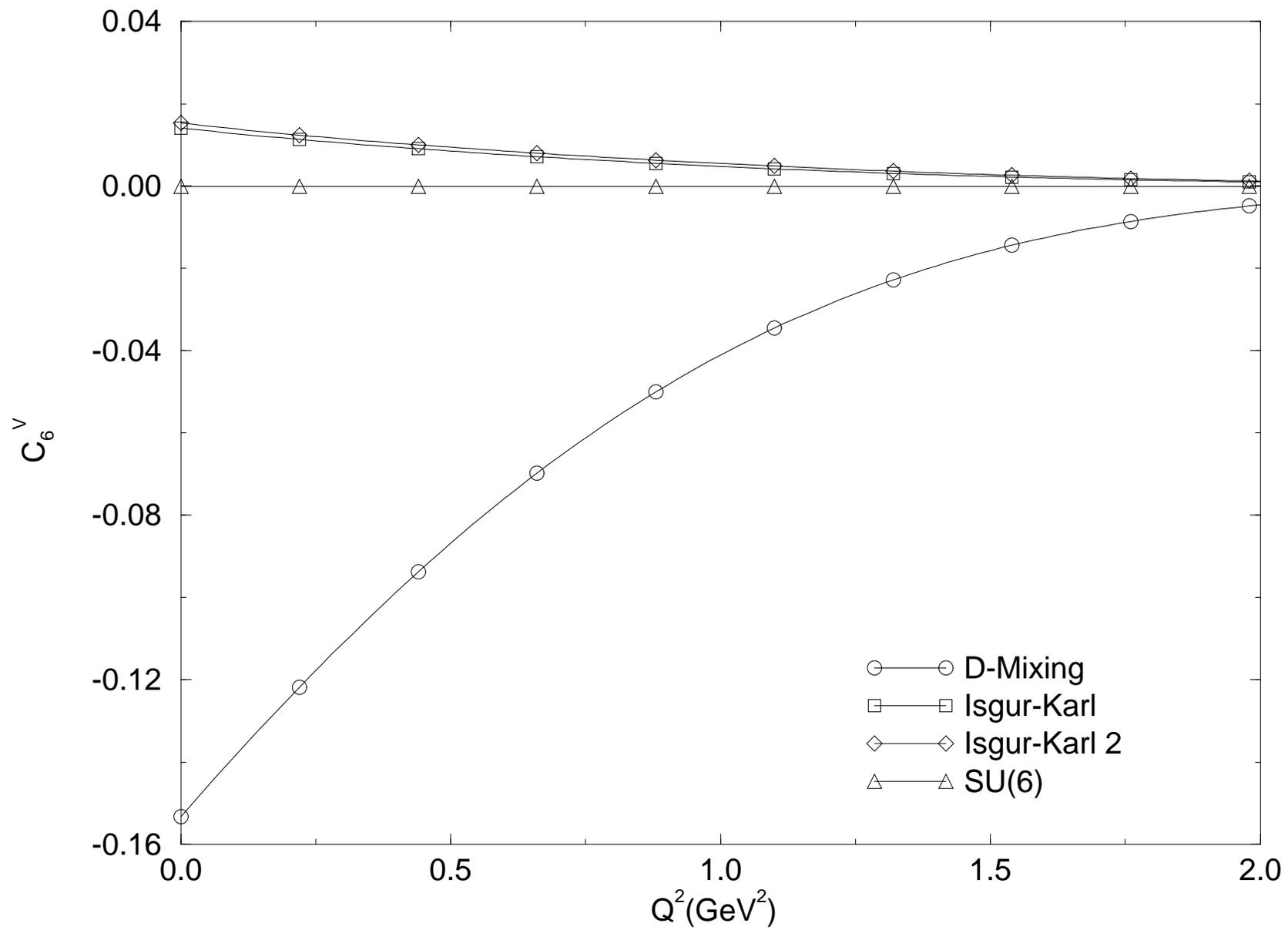
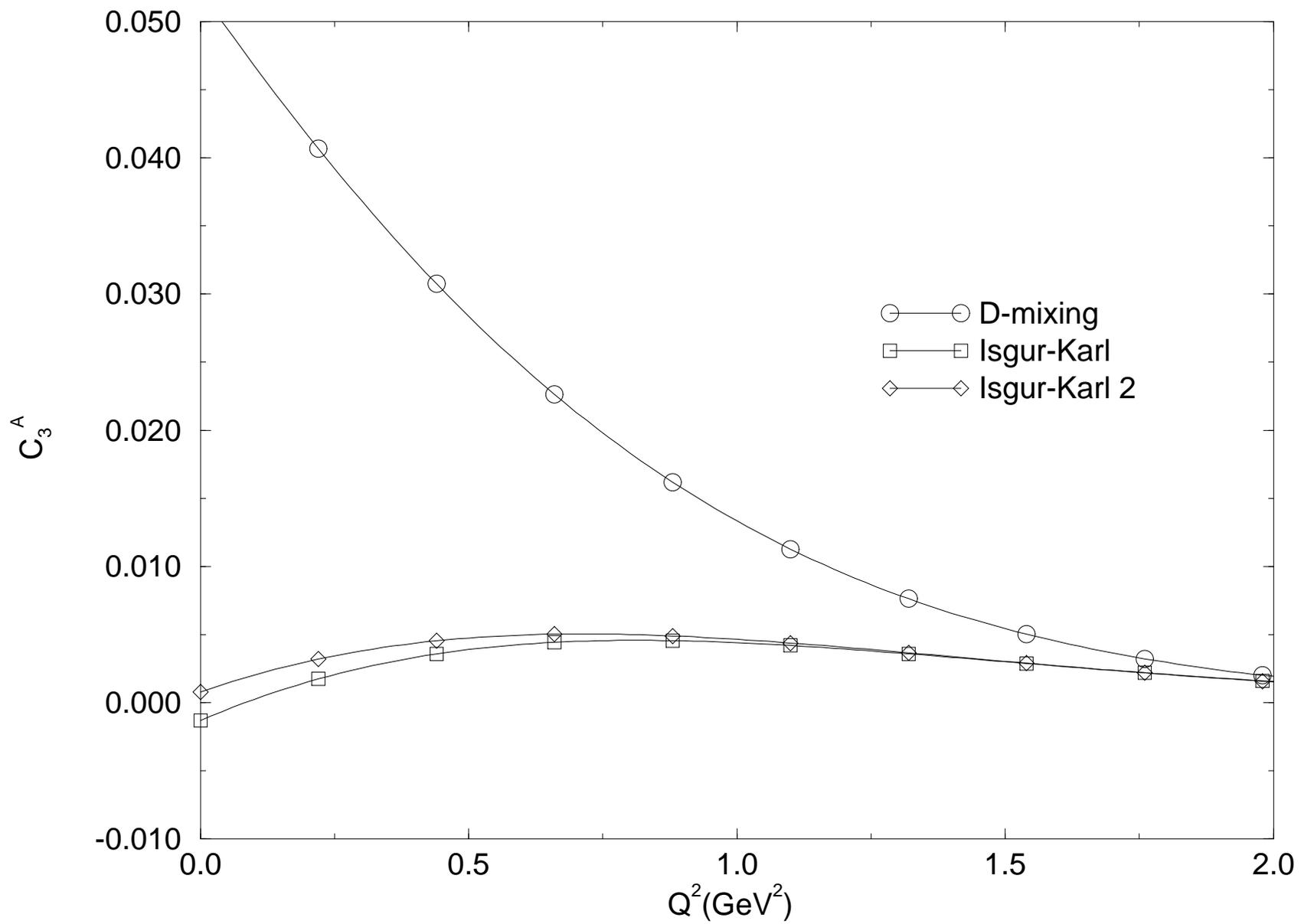
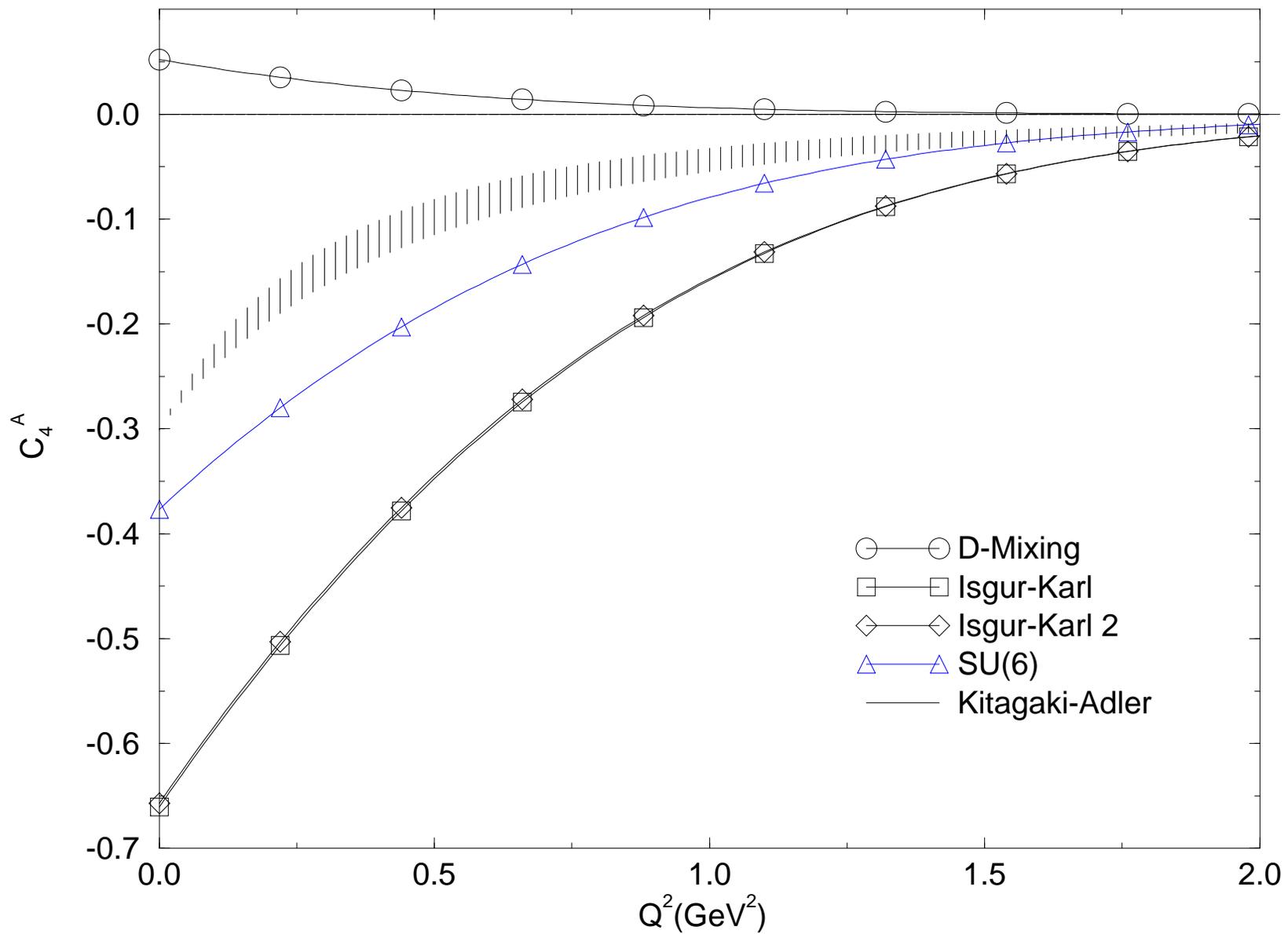
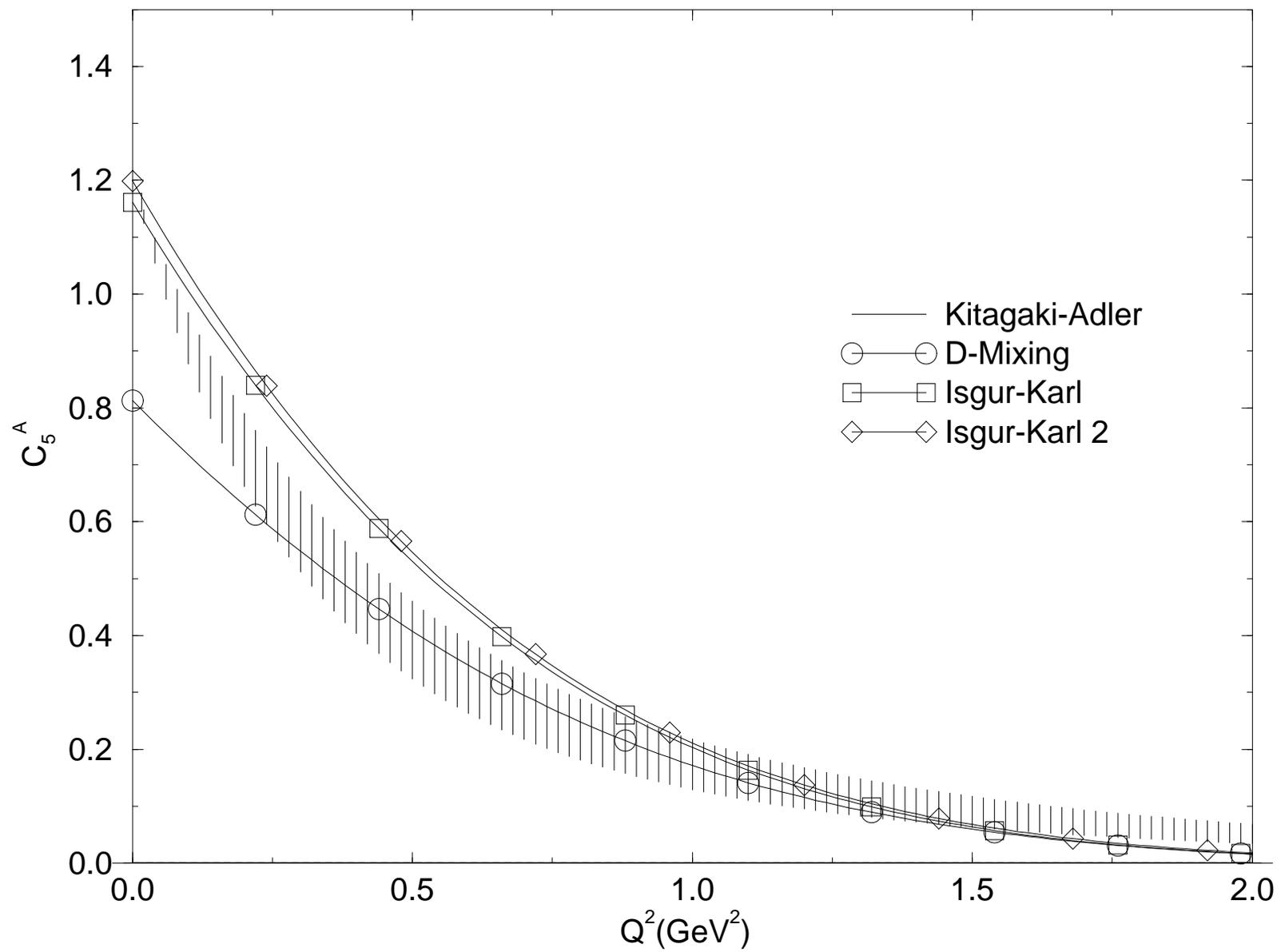
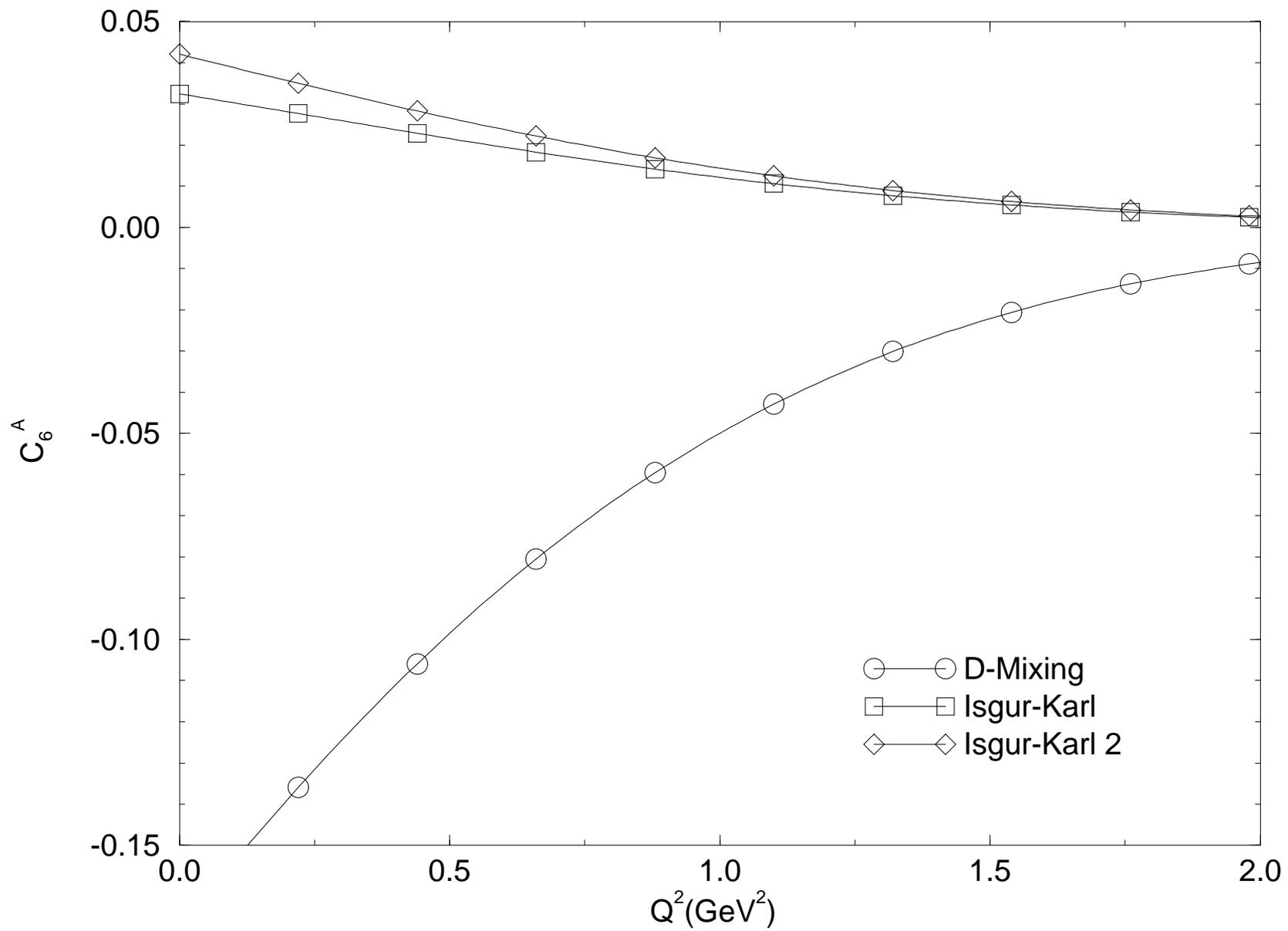
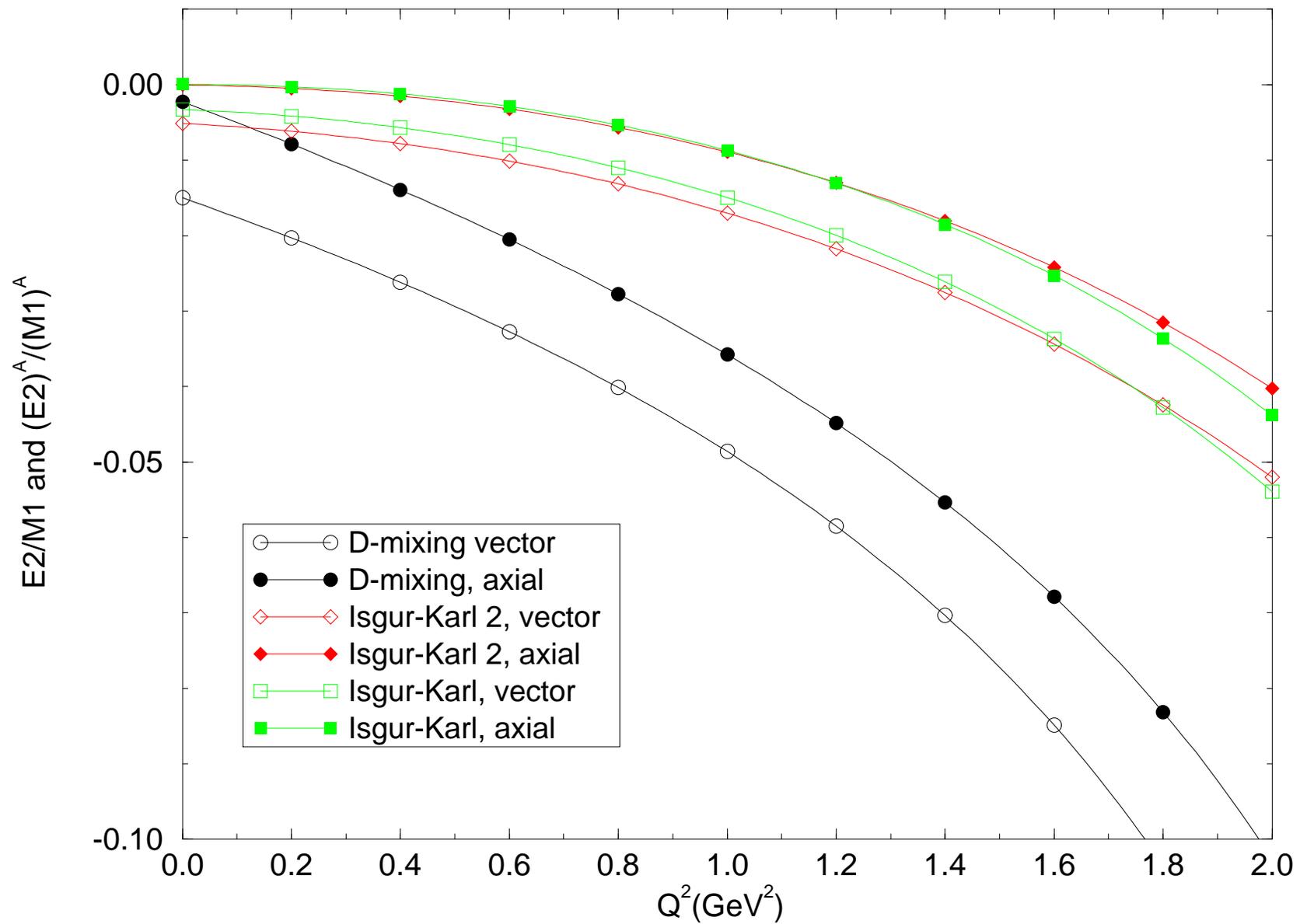
\end{document}